%% file: main.tex
\documentclass[
    journal=largetwo,
    manuscript=preprint,
    year=2025,
]{cup-journal}

\usepackage{amsmath}
\usepackage[nopatch]{microtype}
\usepackage{booktabs}
\usepackage[export]{adjustbox}
\usepackage{float}
\usepackage{graphicx}
\usepackage{array}
\usepackage{tabularx}

\title{Subgroup Performance of a Commercial Digital Breast Tomosynthesis Model for Breast Cancer Detection}

\author{Beatrice Brown-Mulry}
\shortauthor{Brown-Mulry}
\affiliation{HITI Lab, Emory University, Atlanta, GA, USA}
\author{Rohan Satya Isaac}
\affiliation{HITI Lab, Emory University, Atlanta, GA, USA}
\author{Sang Hyup Lee}
\affiliation{Lunit, Seoul, South Korea}

\author{Ambika Seth}
\affiliation{Lunit, Seoul, South Korea}

\author{KyungJee Min}
\affiliation{Lunit, Seoul, South Korea}
\author{Theo Dapamede}
\affiliation{HITI Lab, Emory University, Atlanta, GA, USA}
\author{Frank Li}
\affiliation{HITI Lab, Emory University, Atlanta, GA, USA}
\author{Aawez Mansuri}
\affiliation{HITI Lab, Emory University, Atlanta, GA, USA}
\author{MinJae Woo}
\affiliation{Clemson University, Clemson, SC, USA}
\author{Christian Allison Fauria-Robinson}
\affiliation{Emory University, Atlanta, GA, USA}
\author{Bhavna Paryani}
\affiliation{Emory University, Atlanta, GA, USA}
\author{Judy Wawira Gichoya}
\affiliation{HITI Lab, Emory University, Atlanta, GA, USA}
\author{Hari Trivedi}
\affiliation{HITI Lab, Emory University, Atlanta, GA, USA}


\begin{document}

\begin{abstract}
\input{sections/0_abstract}
\end{abstract}

\section{Background}
\input{sections/1_background}

\section{Methods}
\input{sections/2_methods}

\section{Results}
\input{sections/3_results}

\section{Discussion}
\input{sections/4_discussion}

\section{Conclusion}
\input{sections/5_conclusion}

\section{Acknowledgements}
\input{sections/6_acknowledgements}
\printbibliography
\clearpage
\appendix
\section{Appendix}
\subsection{Supplementary Materials}

\setcounter{figure}{0}
\renewcommand{\thefigure}{\Alph{section}\arabic{figure}}
\setcounter{table}{0}
\renewcommand{\thetable}{\Alph{section}\arabic{table}}

\input{sections/a1_supplement}
\end{document}

%% file: sections/0_abstract.tex
While research has established the potential of AI models for mammography to improve breast cancer screening outcomes, there have not been any detailed subgroup evaluations performed to assess the strengths and weaknesses of commercial models for digital breast tomosynthesis (DBT) imaging. This study presents a granular evaluation of the Lunit INSIGHT DBT model on a large retrospective cohort of 163,449 screening mammography exams from the Emory Breast Imaging Dataset (EMBED). Model performance was evaluated in a binary context with various negative exam types (162,081 exams) compared against screen detected cancers (1,368 exams) as the positive class. The analysis was stratified across demographic, imaging, and pathologic subgroups to identify potential disparities. The model achieved an overall AUC of 0.91 (95\% CI: 0.90–0.92) with a precision of 0.08 (95\% CI: 0.08–0.08), and a recall of 0.73 (95\% CI: 0.71–0.76). Performance was found to be robust across demographics, but cases with non-invasive cancers (AUC: 0.85, 95\% CI: 0.83-0.87), calcifications (AUC: 0.80, 95\% CI: 0.78-0.82), and dense breast tissue (AUC: 0.90, 95\% CI: 0.88-0.91) were associated with significantly lower performance compared to other groups. These results highlight the need for detailed evaluation of model characteristics and vigilance in considering adoption of new tools for clinical deployment.

%% file: sections/1_background.tex
\subsection{Breast Cancer Screening and Digital Breast Tomosynthesis}

Breast cancer is the most common type of cancer in women and causes 42,000 deaths each year in the United States \cite{siegelCancerStatistics20202020}. Screening mammography reduces morbidity and mortality of breast cancers by 38–48\% through early detection of abnormalities and enabling earlier treatment \cite{gotzscheScreeningBreastCancer2013, marmotBenefitsHarmsBreast2013, broedersImpactMammographicScreening2012}. The Breast Imaging – Reporting and Data System (BI-RADS) provides a comprehensive guide of various types of abnormalities that can be seen on mammography, and their level of suspiciousness with respect to cancer \cite{acr2013ACRBIRADS2014}. Despite this, sensitivity and specificity of screening 2D mammography remains between 85-90\% for average-risk women resulting in unnecessary recalls and biopsies, as well as missed cancers \cite{broedersImpactMammographicScreening2012, cunninghamBreastCancerDetection1997, lehmanNationalPerformanceBenchmarks2017}.

Over the past two decades, digital breast tomosynthesis (DBT) has emerged as an advanced imaging modality designed to address the limitations of 2D mammography. DBT provides three-dimensional imaging of the breast and can improve specificity by reducing the masking effect of overlapping tissues \cite{houssamiOverviewEvidenceDigital2013, dhamijaDigitalBreastTomosynthesis2021}. These advantages have led to widespread adoption of the technology since its FDA approval in 2011 \cite{richmanAdoptionDigitalBreast2019}, however research has demonstrated weaknesses in some areas compared to conventional mammography. These include the increased interpretation times of DBT images -- often up to twice as long as 2D mammography--and their potential for reduced visibility of calcifications, a key indicator for certain types of breast cancer \cite{astleyComparisonImageInterpretation2013, zuleyTimeDiagnosisPerformance2010, chenMeasuringReaderFatigue2023}.

\subsection{AI Models for 2D and 3D Mammography}

Over the past several years, multiple deep learning models have been developed both in the research and commercial settings for diagnosis of breast cancer on 2D mammography. A recent meta-analysis demonstrated AI model performance for 2D mammography outperformed radiologists (0.87 vs 0.81, P = .002) in reader studies \cite{yoonStandaloneAIBreast2023}. However, reader studies are difficult to translate to real-world experience due to a manufactured reading environment and upsampling of abnormal cases. In purely retrospective studies, AI model performance for 2D mammography was not shown to be better than radiologists (0.89 vs 0.96, P = .152) \cite{yoonStandaloneAIBreast2023}.  For DBT, three reader studies and one retrospective study demonstrated generally higher AUC for AI compared to radiologists, however with wide confidence intervals \cite{yoonStandaloneAIBreast2023, pintoImpactArtificialIntelligence2021, romero-martinStandAloneUseArtificial2022, shoshanArtificialIntelligenceReducing2022, conantImprovingAccuracyEfficiency2019}. 

\input{tables/label_desc_table}

\begin{figure*}[htb]
\centering
\includegraphics[width=1.0\linewidth]{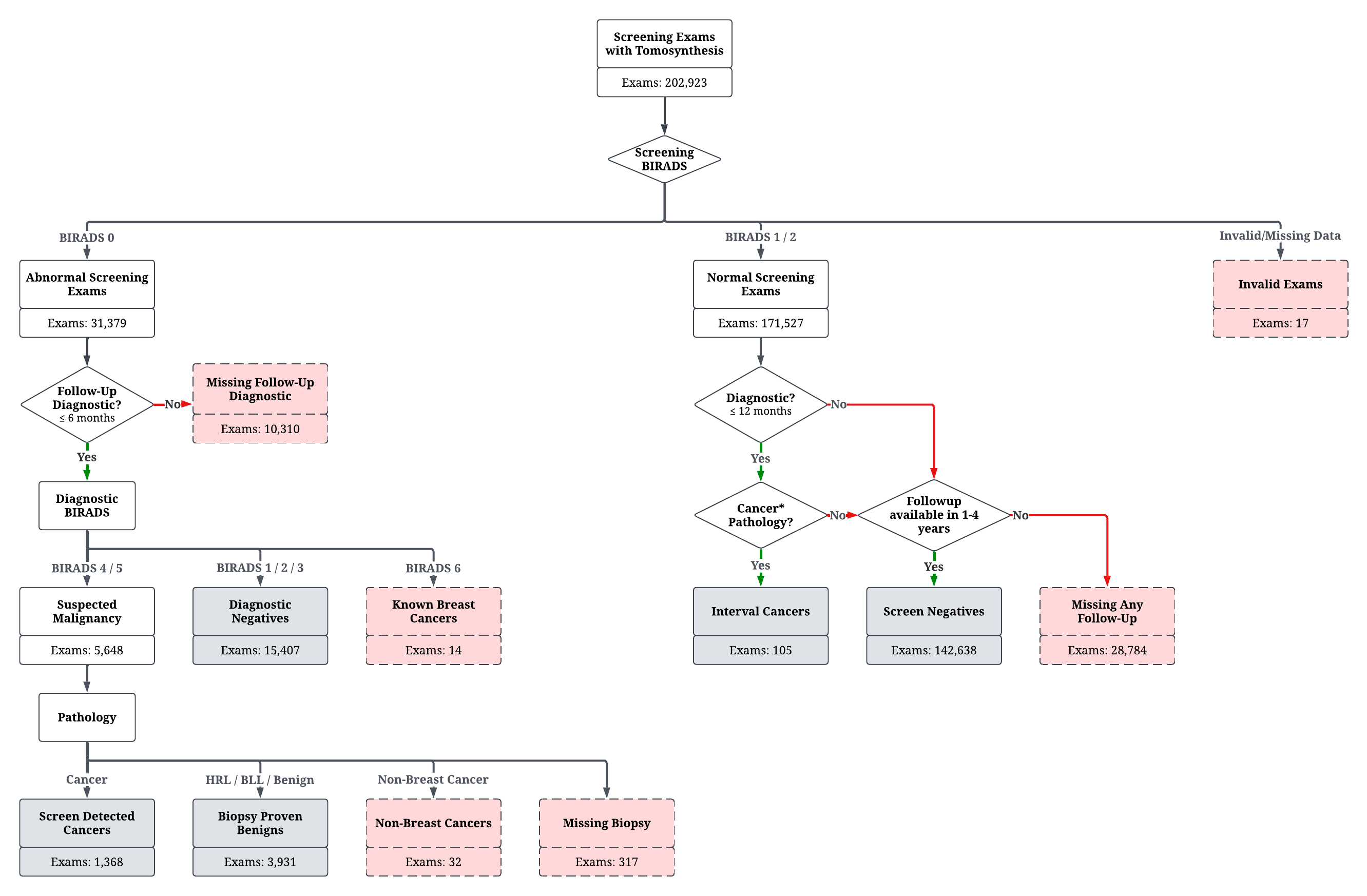}
\caption{Class assignment flowchart. Only the screen-detected cancer label was considered in the positive class, and all others were considered negative with the exception of interval cancers that were considered separately. Abnormal screening exams were divided into confirmed cancers, confirmed benign, and diagnostic negatives. Negative screening exams with at least one follow-up within 1-4 years were considered in the negative class.  Exams that did not meet the criteria for any labels were excluded from further analysis.}
\label{fig:label_flowchart}
\end{figure*}

\subsection{Model Performance across Demographic, Imaging, and Pathological subgroups}

Little work currently exists that studies the performance of breast AI models across various subgroups of patient demographics, imaging characteristics, and pathology subtypes. For AI to gain a foothold in clinical care, clinicians must be able to understand conditions under which models can fail \cite{challenArtificialIntelligenceBias2019, ghassemiFalseHopeCurrent2021, kellyKeyChallengesDelivering2019, rajpurkarAIHealthMedicine2022, purkayasthaFailuresHidingSuccess2021}. Broad interpretation of model performance across all cases can be misleading, since a model that detects the same cancers as a radiologist is not clinically meaningful. Rather, the radiologist and AI model should augment one another with complementary strengths to yield a net positive effect in detection and reduction of false positives \cite{langArtificialIntelligencesupportedScreen2023, hernstromScreeningPerformanceCharacteristics2025, dembrowerArtificialIntelligenceBreast2023}. 

In this study, we evaluate a commercial AI model for breast cancer detection on digital breast tomosynthesis (DBT) on a large retrospective cohort of 163,449 screening mammography exams from 59,358 women \cite{jeongEMoryBrEastImaging2023}, and compare its performance to previously published metrics as well as its FDA validation study. Similar to prior work, we examine model performance across demographics, including race, ethnicity, and age groups. Additionally, performance is considered across clinically meaningful subtypes of cancer (invasive, non-invasive) and subtypes of negative exams (i.e. negative at screening, negative on diagnostic, negative on biopsy). Further, we evaluate performance across finding types (mass, calcification, architectural distortion, and asymmetry) and across specific BI-RADS imaging descriptors.

%% file: tables/label_desc_table.tex
\begin{table}[htb]
\centering
\caption{Description of the exam-level labels considered during evaluation. ‘Screen-detected cancers’ were the only positive class and all others were considered negative for calculation of performance metrics. Interval cancers were excluded from performance metric calculations. Screening exams were considered ‘normal’ if they had exclusively BI-RADS 1 or 2 findings, and ‘abnormal’ if they had at least one finding with a BI-RADS 0 assignment.}
\label{tab:label_desc_table}
\resizebox{\textwidth}{!}{%
\begin{tabular}{lll}
\toprule
Label & Class & Description \\
\midrule
Screen Negative & 0 & \begin{tabular}[c]{@{}l@{}}Normal screening exams with at least \\ one follow-up exam (screening or \\ diagnostic) within 12 months.\end{tabular} \\
Diagnostic Negative & 0 & \begin{tabular}[c]{@{}l@{}}Abnormal screening exams that were \\ assigned a BI-RADS of 1, 2, or 3 on \\ a follow-up diagnostic within 6 months.\end{tabular} \\
Biopsy-Proven Benign & 0 & \begin{tabular}[c]{@{}l@{}}Abnormal screening exams with a biopsy \\ finding on a follow-up diagnostic exam \\ within 6 months indicating a high-risk, \\ borderline, or benign lesion.\end{tabular} \\
Screen-Detected Cancer & 1 & \begin{tabular}[c]{@{}l@{}}Abnormal screening exams with a biopsy \\ finding on a follow-up diagnostic exam \\ within 6 months indicating an invasive \\ or non-invasive cancer.\end{tabular} \\
Interval Cancer & N/a & \begin{tabular}[c]{@{}l@{}}Normal screening exams with a biopsy \\ finding on a follow-up diagnostic exam \\ within 12 months indicating an invasive\\ or non-invasive cancer.\end{tabular} \\
\bottomrule
\end{tabular}
}
\end{table}

%% file: sections/2_methods.tex
\subsection{Model}
The INSIGHT DBT (engine version v.1.3.0; product version v.1.1.2.0) model from Lunit, Inc (Seoul, South Korea) was used for inference in the study. The model received FDA 510k clearance in 2023 and is deployed across 16 countries and 44 sites. The model architecture has 23.4 million trainable parameters and consists of a ResNet34 feature extractor and an aggregation module that outputs an image-level malignancy score and a heatmap for localization. The analysis was conducted using exam-level scores which were derived by selecting the highest predicted image malignancy score for each exam. It was trained on approximately 160,000 exams from 2 countries, and 29\% were cancers. 87\% of the exams were from the USA and 13\% from South Korea. 21\% of the exams were labeled normal, 50\% were benign and 29\% were cancer. Of the exams labeled cancer, 78\% were invasive. 47\% of the exams had a BI-RADS breast density of A or B, rest 53\% had BI-RADS C or D. 4\% of the patients were Black, 9\% Hispanic, 85\% White. 75\% of the exams had patient age between 40 and 79, both inclusive.

\subsection{Data}
This analysis was conducted using digital breast tomosynthesis (DBT) images from the Emory Breast Imaging Dataset (EMBED) \cite{jeongEMoryBrEastImaging2023}. The relevant images were collected from screening exams in four hospitals in the Emory Healthcare network between January 2013 and December 2020. Within EMBED, approximately 72.1 percent of screening exams are 2D + DBT + synthetic 2D, whereas the remainder are 2D only. No exams contain DBT or C-view only. 99.8 percent of exams are from Hologic scanners.

In total, 202,758 exams for 80,299 patients were evaluated by the model, with 58 exams (0.02\%) failing to process. Predicted abnormality scores were aggregated to the exam-level. 163,384 exams for 59,358 patients had unambiguous ground truth data and follow-up available and were considered in the final analysis after excluding 39,374 exams during the label assignment and data handling steps shown in Figure \ref{fig:label_flowchart}.

\subsection{Ground Truthing}
The original radiology report, inclusive of any addendums, was used as ground truth for imaging features and BI-RADS characteristics. Pathology ground truth was obtained from reports and summarized into six categories based on the recommendation of breast pathologists \cite{jeongEMoryBrEastImaging2023}: invasive cancers, non-invasive cancers, high-risk lesions, borderline lesions, benign lesions, and non-breast cancers. Demographics were obtained from the EHR.

\subsection{Label Assignment}
Exams were categorized using a comprehensive set of clinical conditions to determine ground truth labels for evaluation. As all images used in the evaluation were captured at screening, any follow-up diagnostic assessments (recall studies) were matched to their preceding exams. Diagnostic exams for problem evaluations were not considered, except in the case of interval cancers as described below. Descriptions of the five exam labels considered in the analysis are shown in Table \ref{tab:label_desc_table}.

The primary clinical features for label assignment were BI-RADS assessments and pathology results. Because EMBED contains data at the finding level (i.e. a screening exam may have multiple findings in each breast), finding-level assessments were aggregated to the exam-level by selecting the most severe BI-RADS assessment and pathology results from each exam to match the output from the INSIGHT DBT model. 

\begin{figure}[htb]
\centering
\includegraphics[width=0.95\linewidth]{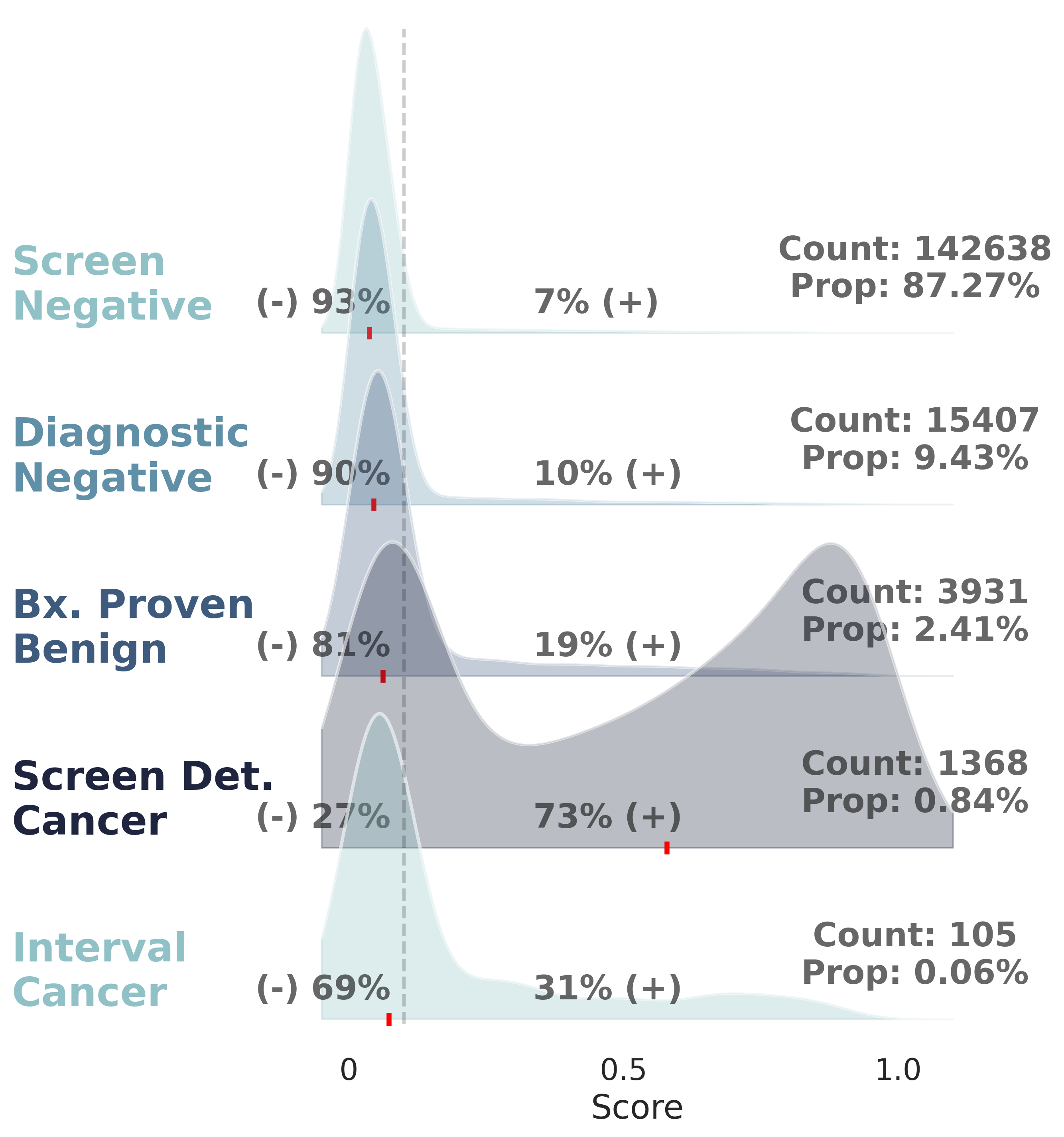}
\caption{Distribution of model predicted scores across outcomes labels: screen negative, diagnostic negative, biopsy proven benign, screen-detected cancer and interval cancer. Model prediction scores range from 0 to 1, with model operating point of 0.1. The chart illustrates the proportion of cases classified as positive or negative based on the set threshold, with shaded areas representing the distribution density.}
\label{fig:outcome_characteristics_with_interval_cancer}
\end{figure}

\subsubsection{Negative Exams}
Three types of ‘negative’ exams were defined: screen negatives (SN), diagnostic negatives (DN), and biopsy-proven benign (B). Screen negatives were defined as screening exams with only BI-RADS 1 or 2 findings that did not fit the interval cancer definition. Diagnostic negatives were defined as screening exams with at least one BI-RADS 0 finding that prompted a diagnostic follow-up which only had BI-RADS 1, 2, or 3 findings. Biopsy-proven benign were defined as abnormal screening exams that prompted a diagnostic follow-up with a BI-RADS 4 or 5 finding which was biopsied and resulted in a high-risk, borderline, or benign lesion. High-risk lesions that were resected and did not result in a cancer were considered negative; those that were resected and yielded a cancer were considered in the positive group (see below). Non-breast cancers like lymphoma and other primary source of cancers were excluded in this evaluation.

\input{tables/demographics_by_label}

\subsubsection{Positive Exams}
Positive exams were classified as screen-detected cancer detected via the normal screening pathway: an abnormal screening exam that prompted a diagnostic follow-up with a BI-RADS 4 or 5 finding followed by an invasive or non-invasive cancer on biopsy or resection. This included high-risk lesions on biopsy that were resected and found to be malignant.

\subsubsection{Interval Cancers}
Interval cancers are defined as a negative screening exam followed by a confirmed cancer diagnosis within 12 months of the screening exam. These were considered as a separate class during evaluation due to the difficulty in categorizing them as either negative or positive. Some interval cancers may be visible on preceding study in retrospect while others were not. However, previous work has shown that AI models were able to detect at least some of these cancers \cite{nanaaAccuracyArtificialIntelligence2024, freemanUseArtificialIntelligence2021, byngAIbasedPreventionInterval2022}. 

\subsubsection{Excluded/Invalid Exams}
Because EMBED relies on human entered data, there are rare circumstances of idiosyncratic or missing data. This includes missing or invalid BI-RADS scores based on exam type, or abnormal screening studies with no diagnostic follow-up or BI-RADS 4 and 5 diagnostic studies with no biopsy results. There are also instances in which a patient leaves the healthcare system and therefore cannot be accurately ground-truthed. Both of these types of exams were excluded from evaluation.

\subsection{Analysis}
\subsubsection{Feature Selection for Subgroup Analysis}
Model performance was assessed across various demographic, imaging, and pathologic subgroups. Demographic subgroups were considered for patient race, ethnicity, and age. The analysis included the patient race categories: Black, White, Asian, unknown or unavailable, and other patient races. The ethnicity groups considered were: Hispanic or Latino, not Hispanic or Latino, and unknown. Patient age at the time of exam was binned into three groups: under-50, 50-to-75, and over-75. 

Imaging features like exam-level presence of masses, asymmetries, architectural distortions, and calcifications were also considered during analysis. Outputs for imaging features stratified by BI-RADS descriptors \cite{acr2013ACRBIRADS2014} are included in the supplement for mass shapes and margins (Figures \ref{fig:mass_shapes_by_outcome_characteristics_boxplot} and \ref{fig:mass_margins_by_outcome_characteristics_boxplot}), calcification morphology and distributions (Figures \ref{fig:calc_distributions_by_outcome_characteristics_boxplot}, \ref{fig:calc_findings_by_outcome_characteristics_suspicious_morphology_boxplot}, and \ref{fig:calc_findings_by_outcome_characteristics_typically_benign_boxplot}), and architectural distortion and asymmetry subtypes (Figures \ref{fig:arch_distortion_by_outcome_characteristics_boxplot} and \ref{fig:asymmetry_by_outcome_characteristics_boxplot}), are included in the supplement.

\input{tables/metrics_table}

Lastly, pathological subgroups examined the final diagnosis resulting from biopsy or subsequent resection. For example, if a patient had an abnormal screening exam followed by diagnostic mammogram and biopsy that resulted in ductal carcinoma in situ (DCIS), followed by resection resulting in invasive ductal carcinoma (IDC), the diagnosis attributed to the screening exam would be IDC. Any subsequent post-treatment diagnoses or recurrences were not considered. Results were stratified by major pathology types: invasive cancers, non-invasive cancers, high-risk lesions, borderline lesions, and benign lesions/normal tissue. Invasive and non-invasive cancers were further decomposed into common subtypes.

\subsubsection{Statistical Evaluation}
Using the ground truth categories determined during label assignment, the model was evaluated as a binary classifier with screen-detected cancers as the positive class (assigned a numeric label of 1) and the negative groups as the negative class (assigned a numeric label of 0). Interval cancers were not considered during metric calculation. Threshold-dependent metrics were evaluated at the commercial operating point for Insight DBT of 0.10. Due to the heavily class imbalance of the data towards then negative class, performance was primarily evaluated with precision, recall, false positive rate (FPR), true negative rate (TNR), false negative rate (FNR), and area under the receiver operating characteristic curve (AUROC).

%% file: tables/demographics_by_label.tex
\begin{table*}[htb]
\centering
\caption{Patient demographics and imaging characteristics by outcome. Positive class was defined as a screen-detected cancer, negative class was defined as a negative screening exam, abnormal screening followed by a negative diagnostic, or an abnormal screening followed by a benign biopsy. Interval cancers were defined as a negative screening exam followed by a cancer within 12 months and were considered separately.}
\label{tab:demographics_by_label}
\resizebox{0.95\textwidth}{!}{%
\begin{tabular}{lrrrrrr}
\toprule
 & Overall & Screen Negative & Diagnostic Negative & Biopsy Proven Benign & Interval Cancer & Screen Detected Cancer\\
\midrule
n & 163449 & 142638 & 15407 & 3931 & 105 & 1368\\
\addlinespace[0.3em]
\multicolumn{7}{l}{\textbf{Race (\%)}}\\
\hspace{1em}Black & 75635 (46.3) & 65851 ( 46.2) & 7019 ( 45.6) & 2069 (52.6) & 43 ( 41.0) & 653 (47.7)\\
\hspace{1em}White & 70763 (43.3) & 62422 ( 43.8) & 6288 ( 40.8) & 1381 (35.1) & 53 ( 50.5) & 619 (45.2)\\
\hspace{1em}Asian & 8602 ( 5.3) & 7497 (  5.3) & 838 (  5.4) & 201 ( 5.1) & 5 (  4.8) & 61 ( 4.5)\\
\hspace{1em}Other & 1472 ( 0.9) & 1191 (  0.8) & 220 (  1.4) & 50 ( 1.3) & 1 (  1.0) & 10 ( 0.7)\\
\hspace{1em}Unknown & 6977 ( 4.3) & 5677 (  4.0) & 1042 (  6.8) & 230 ( 5.9) & 3 (  2.9) & 25 ( 1.8)\\
\addlinespace[0.3em]
\multicolumn{7}{l}{\textbf{Ethnicity (\%)}}\\
\hspace{1em}Not Hispanic or Latino & 135556 (82.9) & 118669 ( 83.2) & 12369 ( 80.3) & 3191 (81.2) & 96 ( 91.4) & 1231 (90.0)\\
\hspace{1em}Hispanic or Latino & 4799 ( 2.9) & 3990 (  2.8) & 657 (  4.3) & 125 ( 3.2) & 0 (  0.0) & 27 ( 2.0)\\
\hspace{1em}Unknown & 23094 (14.1) & 19979 ( 14.0) & 2381 ( 15.5) & 615 (15.6) & 9 (  8.6) & 110 ( 8.0)\\
\addlinespace[0.3em]
\multicolumn{7}{l}{\textbf{Age (\%)}}\\
\hspace{1em}<50 & 38073 (23.3) & 30717 ( 21.5) & 5775 ( 37.5) & 1371 (34.9) & 28 ( 26.7) & 182 (13.3)\\
\hspace{1em}50-75 & 110697 (67.7) & 98571 ( 69.1) & 8693 ( 56.4) & 2364 (60.1) & 71 ( 67.6) & 998 (73.0)\\
\hspace{1em}>=75 & 14679 ( 9.0) & 13350 (  9.4) & 939 (  6.1) & 196 ( 5.0) & 6 (  5.7) & 188 (13.7)\\
\addlinespace[0.3em]
\multicolumn{7}{l}{\textbf{Tissue Density (\%)}}\\
\hspace{1em}A & 17931 (11.0) & 16498 ( 11.6) & 1006 (  6.6) & 327 ( 8.5) & 0 (  0.0) & 100 ( 7.4)\\
\hspace{1em}B & 67228 (41.3) & 59414 ( 41.8) & 5676 ( 37.4) & 1479 (38.4) & 26 ( 25.0) & 633 (46.8)\\
\hspace{1em}C & 68522 (42.1) & 58557 ( 41.2) & 7533 ( 49.7) & 1793 (46.5) & 71 ( 68.3) & 568 (42.0)\\
\hspace{1em}D & 9041 ( 5.6) & 7782 (  5.5) & 948 (  6.3) & 253 ( 6.6) & 7 (  6.7) & 51 ( 3.8)\\
\addlinespace[0.3em]
\multicolumn{7}{l}{\textbf{Screen Detected Pathology (\%)}}\\
\hspace{1em}No Pathology & 158150 (96.8) & 142638 (100.0) & 15407 (100.0) & 0 ( 0.0) & 105 (100.0) & 0 ( 0.0)\\
\hspace{1em}Benign Lesion & 3135 ( 1.9) & 0 (  0.0) & 0 (  0.0) & 3135 (79.8) & 0 (  0.0) & 0 ( 0.0)\\
\hspace{1em}Borderline Lesion & 35 ( 0.0) & 0 (  0.0) & 0 (  0.0) & 35 ( 0.9) & 0 (  0.0) & 0 ( 0.0)\\
\hspace{1em}High Risk Lesion & 761 ( 0.5) & 0 (  0.0) & 0 (  0.0) & 761 (19.4) & 0 (  0.0) & 0 ( 0.0)\\
\hspace{1em}Invasive Cancer & 914 ( 0.6) & 0 (  0.0) & 0 (  0.0) & 0 ( 0.0) & 0 (  0.0) & 914 (66.8)\\
\hspace{1em}Noninvasive Cancer & 454 ( 0.3) & 0 (  0.0) & 0 (  0.0) & 0 ( 0.0) & 0 (  0.0) & 454 (33.2)\\
\addlinespace[0.3em]
\multicolumn{7}{l}{\textbf{Finding Characteristics (\%)}}\\
\hspace{1em}Mass (\%) & 5987 ( 3.7) & 2269 (  1.6) & 2775 ( 18.0) & 707 (18.0) & 0 (  0.0) & 236 (17.3)\\
\hspace{1em}Asymmetry (\%) & 11528 ( 7.1) & 753 (  0.5) & 8731 ( 56.7) & 1503 (38.2) & 3 (  2.9) & 538 (39.3)\\
\hspace{1em}Architectural Distortion (\%) & 2023 ( 1.2) & 38 (  0.0) & 1419 (  9.2) & 352 ( 9.0) & 0 (  0.0) & 214 (15.6)\\
\hspace{1em}Calcification (\%) & 6081 ( 3.7) & 1507 (  1.1) & 2204 ( 14.3) & 1741 (44.3) & 2 (  1.9) & 627 (45.8)\\
\bottomrule
\end{tabular}%
}
\end{table*}

%% file: tables/metrics_table.tex
\begin{table*}[htb]
\centering
\caption{Model performance overall and stratified by demographic, imaging, and pathologic subgroups. Summarize the salient findings. Metrics are presented with 95\% confidence intervals. Sensitivity, specificity, and AUC values are reported with 95\% confidence intervals as provided by the FDA. Additional metrics (precision, FPR, and FNR) were derived based on the reported prevalence and are presented without confidence intervals.}
\label{tab:metrics_table}
\resizebox{0.95\textwidth}{!}{%
\begin{tabular}{lrrrrrrrr}
\toprule
Group Value & Total Negatives & Total Positives & Precision & Recall & FPR & TNR & FNR & AUC\\
\midrule
FDA & \multicolumn{1}{c}{-} & \multicolumn{1}{c}{-} & \multicolumn{1}{r}{0.84 (\ \ \ \ \ \ N/A\ \ \ \ \ \ )} & 0.86 (0.84, 0.88) & \multicolumn{1}{r}{0.16 (\ \ \ \ \ \ N/A\ \ \ \ \ \ )} & 0.84 (0.82, 0.86) & \multicolumn{1}{r}{0.14 (\ \ \ \ \ \ N/A\ \ \ \ \ \ )} & 0.93 (0.92, 0.94)\\
Overall & 161976 (99.16\%) & 1368 (0.84\%) & 0.08 (0.08, 0.08) & 0.73 (0.71, 0.76) & 0.07 (0.07, 0.07) & 0.93 (0.93, 0.93) & 0.27 (0.24, 0.29) & 0.91 (0.90, 0.92)\\
\addlinespace[0.3em]
\multicolumn{9}{l}{\textbf{Race}}\\
\hspace{1em}Black & 74939 (99.14\%) & 653 (0.86\%) & 0.08 (0.07, 0.09) & 0.71 (0.67, 0.74) & 0.07 (0.07, 0.07) & 0.93 (0.93, 0.93) & 0.29 (0.26, 0.33) & 0.91 (0.89, 0.92)\\
\hspace{1em}White & 70091 (99.12\%) & 619 (0.88\%) & 0.09 (0.08, 0.09) & 0.77 (0.73, 0.80) & 0.07 (0.07, 0.07) & 0.93 (0.93, 0.93) & 0.23 (0.20, 0.27) & 0.92 (0.91, 0.94)\\
\hspace{1em}Asian & 8536 (99.29\%) & 61 (0.71\%) & 0.06 (0.04, 0.07) & 0.69 (0.57, 0.81) & 0.08 (0.08, 0.09) & 0.92 (0.91, 0.92) & 0.31 (0.19, 0.43) & 0.88 (0.82, 0.92)\\
\hspace{1em}Unknown & 6949 (99.64\%) & 25 (0.36\%) & 0.04 (0.03, 0.06) & 0.80 (0.63, 0.94) & 0.06 (0.06, 0.07) & 0.94 (0.93, 0.94) & 0.20 (0.06, 0.37) & 0.94 (0.88, 0.98)\\
\hspace{1em}Other & 1461 (99.32\%) & 10 (0.68\%) & 0.07 (0.02, 0.13) & 0.60 (0.29, 0.91) & 0.06 (0.04, 0.07) & 0.94 (0.93, 0.96) & 0.40 (0.09, 0.71) & 0.86 (0.73, 0.98)\\
\addlinespace[0.3em]
\multicolumn{9}{l}{\textbf{Ethnicity}}\\
\hspace{1em}Not Hispanic or Latino & 134229 (99.09\%) & 1231 (0.91\%) & 0.09 (0.08, 0.09) & 0.73 (0.71, 0.76) & 0.07 (0.07, 0.07) & 0.93 (0.93, 0.93) & 0.27 (0.24, 0.29) & 0.91 (0.90, 0.92)\\
\hspace{1em}Hispanic or Latino & 4772 (99.44\%) & 27 (0.56\%) & 0.05 (0.03, 0.08) & 0.70 (0.52, 0.87) & 0.07 (0.06, 0.08) & 0.93 (0.92, 0.94) & 0.30 (0.13, 0.48) & 0.87 (0.78, 0.95)\\
\hspace{1em}Unknown & 22975 (99.52\%) & 110 (0.48\%) & 0.05 (0.04, 0.06) & 0.75 (0.66, 0.83) & 0.07 (0.06, 0.07) & 0.93 (0.93, 0.94) & 0.25 (0.17, 0.34) & 0.93 (0.90, 0.95)\\
\addlinespace[0.3em]
\multicolumn{9}{l}{\textbf{Tissue Density}}\\
\hspace{1em}A & 17831 (99.44\%) & 100 (0.56\%) & 0.08 (0.06, 0.10) & 0.73 (0.64, 0.82) & 0.05 (0.04, 0.05) & 0.95 (0.95, 0.96) & 0.27 (0.18, 0.36) & 0.95 (0.92, 0.97)\\
\hspace{1em}B & 66569 (99.06\%) & 633 (0.94\%) & 0.09 (0.08, 0.10) & 0.77 (0.74, 0.80) & 0.08 (0.07, 0.08) & 0.92 (0.92, 0.93) & 0.23 (0.20, 0.26) & 0.92 (0.91, 0.93)\\
\hspace{1em}C & 67883 (99.17\%) & 568 (0.83\%) & 0.07 (0.07, 0.08) & 0.71 (0.67, 0.74) & 0.07 (0.07, 0.08) & 0.93 (0.92, 0.93) & 0.29 (0.26, 0.33) & 0.90 (0.88, 0.91)\\
\hspace{1em}D & 8983 (99.44\%) & 51 (0.56\%) & 0.05 (0.03, 0.07) & 0.62 (0.49, 0.76) & 0.07 (0.06, 0.07) & 0.93 (0.93, 0.94) & 0.38 (0.24, 0.51) & 0.87 (0.82, 0.92)\\
\addlinespace[0.3em]
\multicolumn{9}{l}{\textbf{Age}}\\
\hspace{1em}<50 & 37863 (99.52\%) & 182 (0.48\%) & 0.07 (0.06, 0.08) & 0.71 (0.64, 0.77) & 0.05 (0.05, 0.05) & 0.95 (0.95, 0.95) & 0.29 (0.23, 0.36) & 0.93 (0.91, 0.95)\\
\hspace{1em}50-75 & 109628 (99.10\%) & 998 (0.90\%) & 0.08 (0.08, 0.09) & 0.72 (0.69, 0.75) & 0.07 (0.07, 0.07) & 0.93 (0.93, 0.93) & 0.28 (0.25, 0.31) & 0.91 (0.90, 0.92)\\
\hspace{1em}>=75 & 14485 (98.72\%) & 188 (1.28\%) & 0.08 (0.07, 0.09) & 0.81 (0.76, 0.87) & 0.12 (0.11, 0.12) & 0.88 (0.88, 0.89) & 0.19 (0.13, 0.24) & 0.91 (0.89, 0.94)\\
\addlinespace[0.3em]
\multicolumn{9}{l}{\textbf{Screen Detected Cancer Type}}\\
\hspace{1em}Invasive Cancer & 161976 (99.44\%) & 914 (0.56\%) & 0.06 (0.06, 0.07) & 0.83 (0.80, 0.85) & 0.07 (0.07, 0.07) & 0.93 (0.93, 0.93) & 0.17 (0.15, 0.20) & 0.94 (0.94, 0.95)\\
\hspace{1em}Non-Invasive Cancer & 161976 (99.72\%) & 454 (0.28\%) & 0.02 (0.02, 0.02) & 0.55 (0.50, 0.60) & 0.07 (0.07, 0.07) & 0.93 (0.93, 0.93) & 0.45 (0.40, 0.50) & 0.85 (0.83, 0.87)\\
\addlinespace[0.3em]
\multicolumn{9}{l}{\textbf{Finding Characteristics}}\\
\hspace{1em}Mass & 5751 (96.06\%) & 236 (3.94\%) & 0.21 (0.18, 0.24) & 0.85 (0.80, 0.89) & 0.13 (0.12, 0.14) & 0.87 (0.86, 0.88) & 0.15 (0.11, 0.20) & 0.93 (0.91, 0.95)\\
\hspace{1em}Asymmetry & 10987 (95.33\%) & 538 (4.67\%) & 0.28 (0.26, 0.30) & 0.78 (0.75, 0.82) & 0.10 (0.09, 0.10) & 0.90 (0.90, 0.91) & 0.22 (0.18, 0.25) & 0.92 (0.90, 0.93)\\
\hspace{1em}Architectural Distortion & 1809 (89.42\%) & 214 (10.58\%) & 0.34 (0.30, 0.38) & 0.83 (0.78, 0.88) & 0.19 (0.17, 0.21) & 0.81 (0.79, 0.83) & 0.17 (0.12, 0.22) & 0.90 (0.88, 0.92)\\
\hspace{1em}Calcification & 5452 (89.69\%) & 627 (10.31\%) & 0.29 (0.27, 0.31) & 0.66 (0.62, 0.69) & 0.18 (0.17, 0.19) & 0.82 (0.81, 0.83) & 0.34 (0.31, 0.38) & 0.80 (0.78, 0.82)\\
\bottomrule
\end{tabular}%
}
\end{table*}

%% file: sections/3_results.tex
A total of 161,976 negative exams and 1,368 positive exams were included in the final evaluation of the model (Figure \ref{fig:label_flowchart}). A summary of distribution is shown in Table \ref{tab:demographics_by_label}. The majority of exams were for Black (75,635, 46.3\%) and White patients (70,763, 43.3\%), with 653 (47.7\%) and 619 (45.2\%) screen-detected cancers, respectively. Non-Hispanic or Latino patients (135,556, 82.9\%) represented the major ethnicity.  Breast density was distributed as A (17,931, 11.0\%), B (67,228, 41.3\%), C (68,522, 42.1\%), and D (9041, 5.6\%).

\subsection{Model Performance}
The model achieved an overall AUC of 0.91 (95\% CI: 0.90–0.92) in distinguishing between screen-detected cancers and negative exams, with a recall of 0.73, FPR of 0.07, TNR of 0.93, and FNR of 0.27 (Table \ref{tab:metrics_table}). Performance was not statistically different across Race, Ethnicity, or Age. In addition, model performance was evaluated across multiple subgroups including screening outcome, breast density, imaging features, and pathology subtype.

\subsection{Model Trends by Subgroup}
\subsubsection{Outcome subgroups}
We observed that model prediction scores trended slightly higher between screen negative, diagnostic negative, and confirmed benign exams, with decreasing percentages of exams being classified as normal according to the operating point of 0.1 (Figure \ref{fig:outcome_characteristics_with_interval_cancer}). The model classified 93\% (133,341/142,638) of screening studies as negative, 90\% (13,930/15,407) of diagnostic negative exams as negative, and 81\% (3,193/3,931) biopsy proven benign exams as negative. The model classified 73\% (1,002/ 1,368) of screen-detected cancers as positive. Finally, the model classified 31\% (33/105) of interval cancers as positive.

\begin{figure}[htb]
\centering
\includegraphics[width=0.95\linewidth]{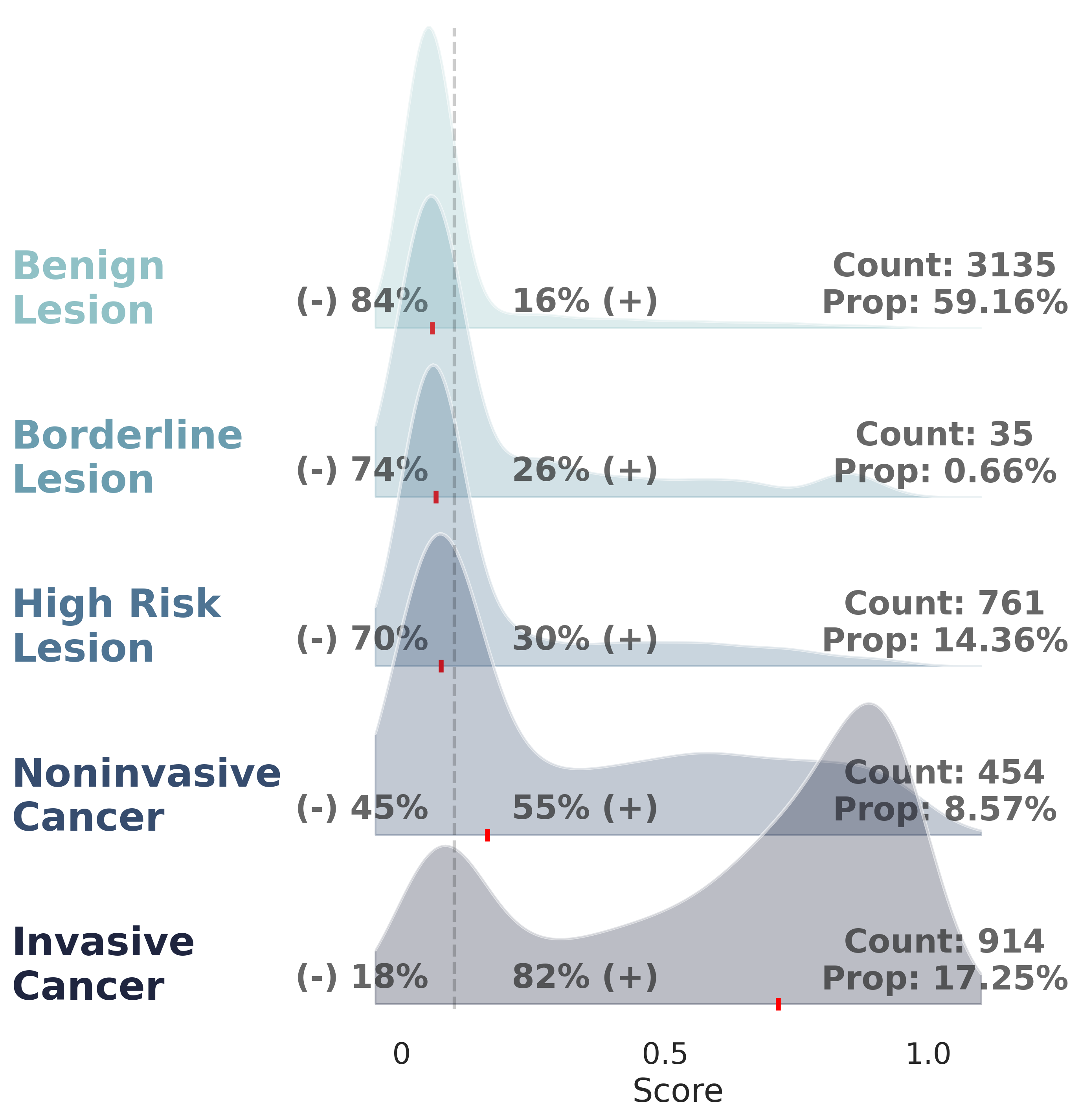}
\caption{Distribution of model predicted scores for pathology subtypes. Each category represents the most severe pathology identified for a given exam. Model prediction scores range from 0 to 1, with model operating point of 0.1. The chart illustrates the proportion of cases classified as positive or negative based on the set threshold, with shaded areas representing the distribution density.}
\label{fig:outcomes_by_pathology}
\end{figure}

\subsubsection{Pathological Subgroups}
For binary classification, we considered invasive and non-invasive cancers as positive, and all other cases as negative (screen negative, diagnostic negative, benign, borderline, and high-risk lesions). We observed that model prediction scores trended higher as pathology subtypes became more abnormal or at higher risk of becoming cancer. At the operating point of 0.1, the model correctly classified 84\% (2,635/3,135) of screening studies with a benign lesion, 74\% (26/35) of studies with a borderline lesion and 70\% (532/761) of studies with a high-risk lesion as negative. For screening studies with non-invasive and invasive cancer outcomes, the model correctly identified 55\% (248/454) of non-invasive cancers and 82\% (754/914) of invasive cancers (Figure \ref{fig:outcomes_by_pathology}).

\begin{figure*}[htbp]
\centering
\includegraphics[width=0.95\linewidth]{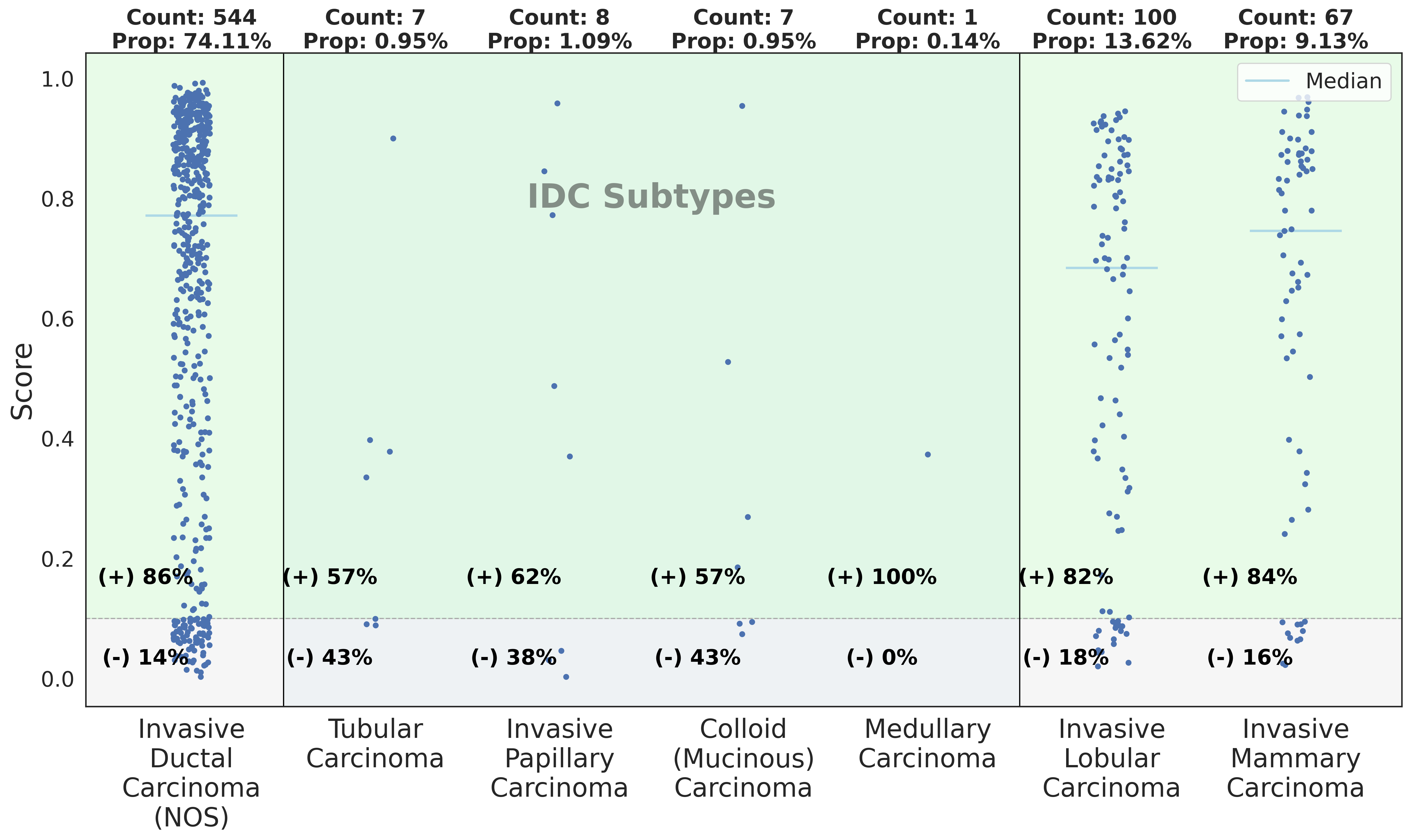}
\caption{Distribution of model predicted scores for various invasive cancer subtypes. The model operating point of 0.1 is marked by a horizontal line with the percentage of exams above and below the threshold indicated per pathology. IDC has many subtypes with varying imaging features and prognosis, and are represented by the adjacent panel. Model performance for invasive cancers is generally good, with lower proportion of positive cases within certain subtypes although low number of samples precludes any conclusions regarding subtypes.}
\label{fig:invasive_cancer_pathology_outcomes}
\end{figure*}

Performance was generally good across invasive cancer subtypes but trended lower for certain rare subtypes. Invasive Ductal Carcinoma, Not Otherwise Specified (IDC NOS) represented the majority of cases (544/734), and had 86\% (468/544) of cases classified as positive. Invasive Lobular Carcinoma (ILC) was classified correctly in 82\% (82/100) of cases and Invasive Mammary Carcinoma (IMC) in 84\% (56/67) of cases (Figure \ref{fig:invasive_cancer_pathology_outcomes}). Note that the aforementioned counts consider the number of invasive pathologies present, which differs from the total number of screen detected invasive cancers exams as some contain multiple invasive pathologies.

\begin{figure*}[htbp]
\centering
\includegraphics[width=0.95\linewidth]{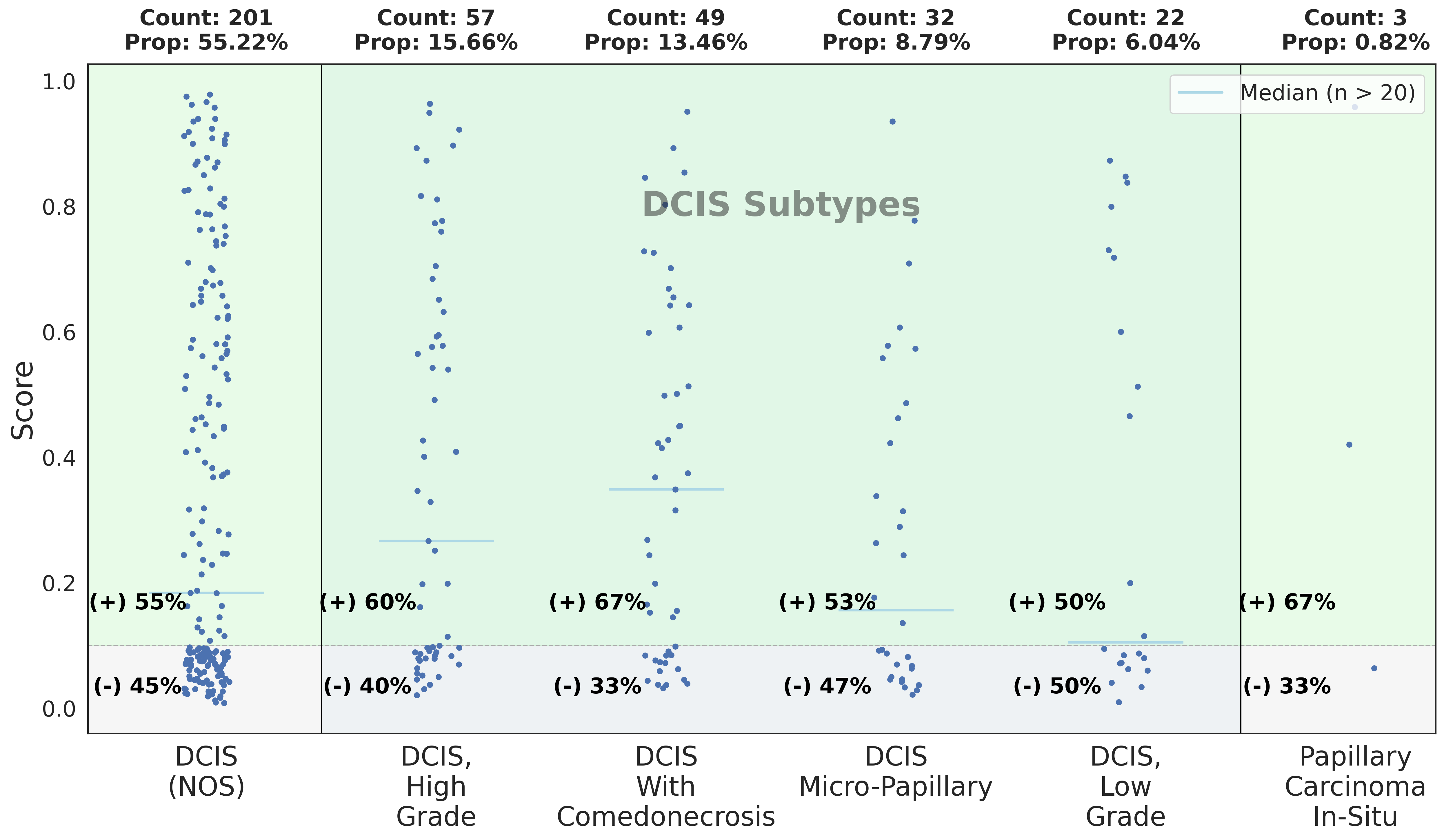}
\caption{Distribution of model predicted scores for various non-invasive cancer subtypes. The model operating point of 0.1 is marked by a horizontal line with the percentage of exams above and below the threshold indicated per pathology. DCIS grades and subtypes are represented in the adjacent panel. Model performance for non-invasive cancers was significantly lower for all non-invasive cancers compared to invasive cancers.}
\label{fig:noninvasive_cancer_pathology_outcomes}
\end{figure*}

Model performance was significantly worse for non-invasive cancers (AUC 0.85, 95\% CI: 0.83-0.87; Recall 0.55, 95\% CI: 0.50-0.60; n=454; p<0.05) compared to invasive cancers (AUC 0.94, 95\% CI: 0.94–0.95; Recall 0.83, 95\% CI: 0.80-0.85;  n=914; p<=0.05) across all non-invasive sub-types and grades (Figure \ref{fig:noninvasive_cancer_pathology_outcomes}).  Ductal Carcinoma In Situ, Not Otherwise Specified (DCIS NOS) represented the majority of cases (201/364) and had 55\% (111/201) of cases classified as positive. Note that the aforementioned counts consider the number of non-invasive pathologies present which will differ than the total number of screen detected non-invasive cancers, as some exams contained multiple non-invasive pathologies.

\subsubsection{By Imaging Feature}
\begin{figure*}[htb]
\centering
\includegraphics[width=0.9\linewidth]{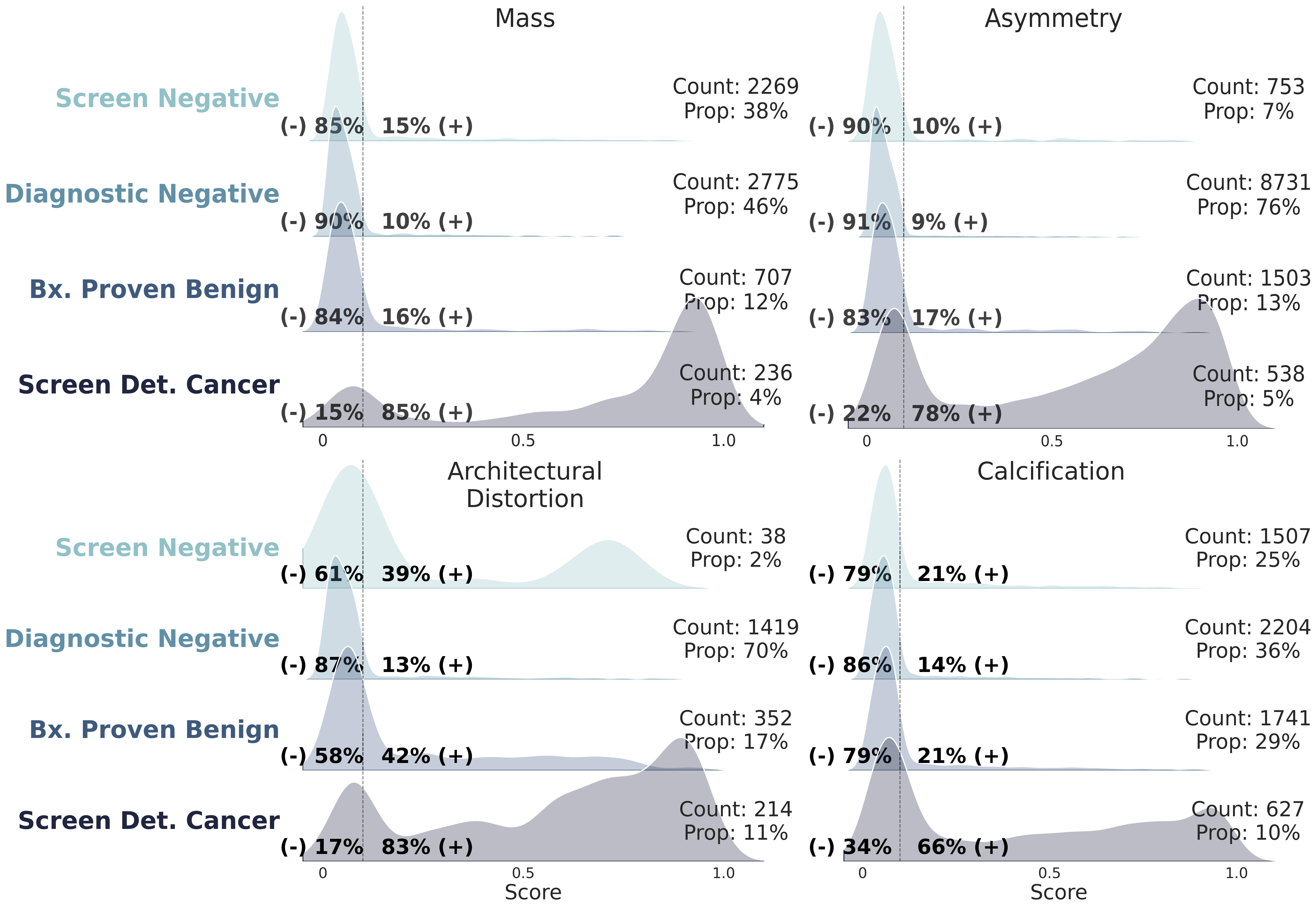}
\caption{Distribution of model prediction score for exams containing various imaging features, structured by clinical outcome. For mass, asymmetry, and calcification, we observe generally that images containing these features are correctly predicted as negative in 60-81\% of exams. However, architectural distortions are much less likely to be predicted as malignant regardless of pathologic outcome. Interestingly, architectural distortions from screen negative exams had a strong bimodal distribution of high and low scares, yielding many false positives.}
\label{fig:binary_findings_by_outcome_characteristics}
\end{figure*}

Model performance varied based on the presence of specific findings in screening studies (Figure \ref{fig:binary_findings_by_outcome_characteristics}). Because exams could contain multiple findings and model results are returned at the exam level, our results are aggregated for the presence of any imaging finding at the study level. For example, if an exam has both a mass and calcification, the model output is attributed to both the mass and calcification subgroups. 82\% (17,002) of abnormal screening exams had only one BI-RADS 0 lesion, while 18\% of abnormal exams (3,704) contained more than one. For exams containing a mass, the model correctly classified 85\% (1918 of 2269) of cases in the screen negative group, 90\% (2488 of 2775) in the diagnostic negative group, 84\% (593 of 707) in the confirmed benign group, and 85\% (200 of 236) in the screen-detected cancer group. For studies with asymmetry, the model correctly classified 90\% (674 of 753) in the screen negative group, 91\% (7979 of 8731) in the diagnostic negative group, 83\% (1248 of 1503) in the confirmed benign group, and 78\% (420 of 538) in the screen-detected cancer group. For studies with architectural distortion, the model correctly classified 61\% (23 of 38) of cases in the screen negative group, 87\% (1235 of 1419) in the diagnostic negative group, 58\% (204 of 352) in the confirmed benign group, and 83\% (178 of 214) in the screen-detected cancer group. For studies with calcification, the model correctly classified 79\% (1191 of 1507) of cases in the screen negative group, 86\% (1887 of 2204) in the diagnostic negative group, 79\% (1378 of 1741) in the confirmed benign group, and 66\% (412 of 627) in the screen-detected cancer group. Outputs for imaging features were further stratified by BI-RADS descriptors such as mass shapes and margins (Figures \ref{fig:mass_shapes_by_outcome_characteristics_boxplot} and \ref{fig:mass_margins_by_outcome_characteristics_boxplot}), calcification morphology and distribution (Figures \ref{fig:calc_distributions_by_outcome_characteristics_boxplot}, \ref{fig:calc_findings_by_outcome_characteristics_suspicious_morphology_boxplot}, and \ref{fig:calc_findings_by_outcome_characteristics_typically_benign_boxplot}), and architectural distortion and asymmetry subtypes (Figures \ref{fig:arch_distortion_by_outcome_characteristics_boxplot} and \ref{fig:asymmetry_by_outcome_characteristics_boxplot}). 

Qualitatively, exams with masses showed the best separation between model prediction scores of positive and negative exams whereas exams with architectural distortion showed the least separation. Examples of true positive, false positive, and false negative model predictions by lesion type are shown in Figure \ref{fig:imaging_features_grid}. 

\input{tables/imaging_features_grid_table_TESTING}

%% file: tables/imaging_features_grid_table_TESTING.tex
\newcolumntype{C}{>{\centering\arraybackslash} m{0.27\textwidth} }
\begin{figure}[htb]
    \centering
    \setlength{\tabcolsep}{1.5pt} 
    \resizebox{\textwidth}{!}{%
    \begin{tabular}{m{0.19\textwidth} CCC} 
        & \textbf{TP} & \textbf{FP} & \textbf{FN} \\
        
        \textbf{Mass} & 
        \includegraphics[width=0.27\textwidth]{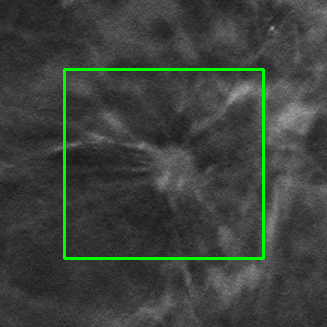} &
        \includegraphics[width=0.27\textwidth]{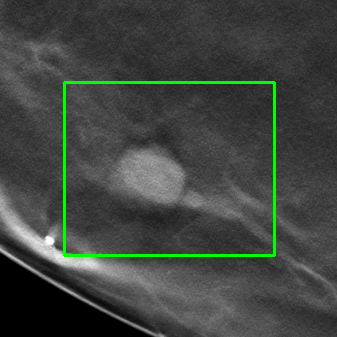} &
        \includegraphics[width=0.27\textwidth]{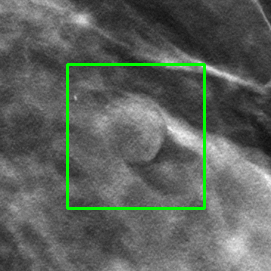} \\
        
        \textbf{Asymmetry} & 
        \includegraphics[width=0.27\textwidth]{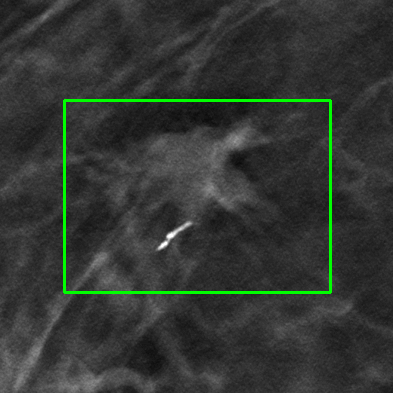} &
        \includegraphics[width=0.27\textwidth]{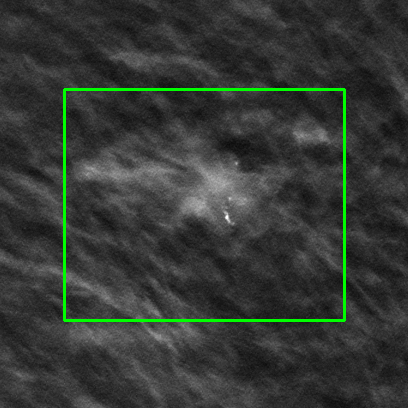} &
        \includegraphics[width=0.27\textwidth]{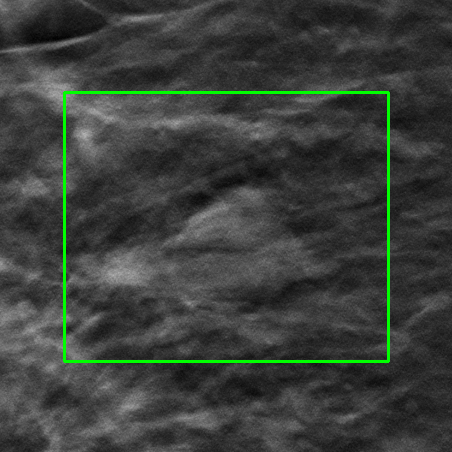} \\
        
        \begin{tabular}[c]{@{}l@{}}\textbf{Architectural}\\\textbf{Distortion}\end{tabular} & 
        \includegraphics[width=0.27\textwidth]{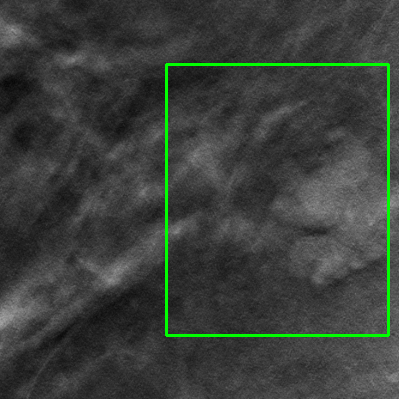} &
        \includegraphics[width=0.27\textwidth]{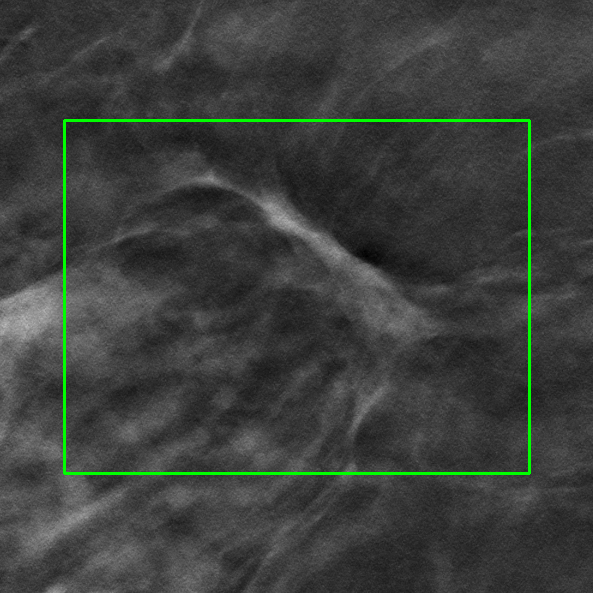} &
        \includegraphics[width=0.27\textwidth]{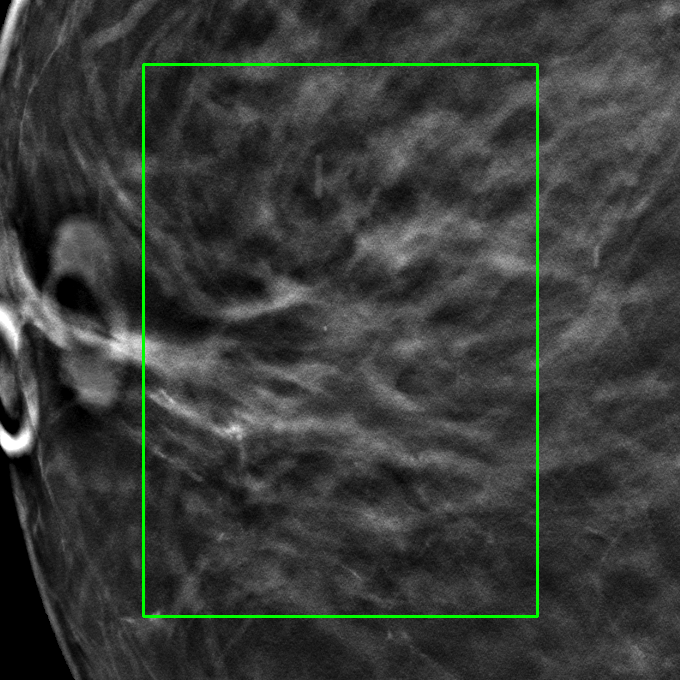} \\
        
        \textbf{Calcification} & 
        \includegraphics[width=0.27\textwidth]{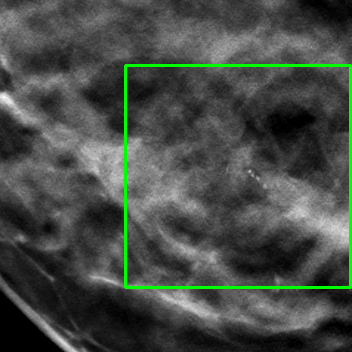} &
        \includegraphics[width=0.27\textwidth]{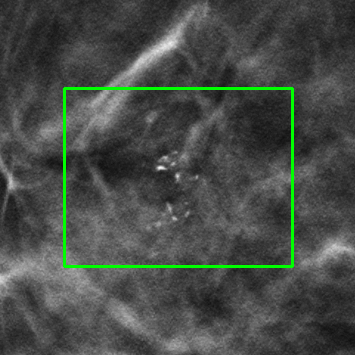} &
        \includegraphics[width=0.27\textwidth]{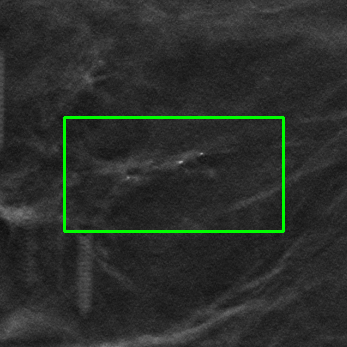} \\
    \end{tabular}
    }
    \caption{Examples of model predictions across imaging features of mass, asymmetry, architectural distortion, and calcifications. True positive lesions are from exams containing cancer that were correctly identified by the model. False positive lesions are from cases in which the model predicted cancer for exams that were either deemed negative at diagnostic imaging or benign following biopsy. False negative lesions are from exams that were malignant on biopsy, however predicted as negative by the model. Note that the model returns only breast-level and exam-level predictions and not regions of interest and these lesions are highlighted (green box) for demonstration purposes. }
    \label{fig:imaging_features_grid}
\end{figure}

%% file: sections/4_discussion.tex
This study represents the first performance evaluation of a commercial AI model for breast cancer detection across demographic, pathologic, and imaging subtypes.  Breast cancer, like most diseases, is diverse in its presentation, imaging features, demographic distribution, and severity. Artificially binarizing disease into positive and negative classes without considering clinically meaningful subtypes masks true model performance, as no two cancers are alike. When considering permutations of race, ethnicity, breast density, cancer subtype, and four main imaging findings, our test set contained 1368 cancers with 215 unique permutations of findings. If adding various subtypes of masses and calcifications, nearly every cancer would be unique. Because of the large variety of presentations, lack of understanding of a models’ strengths and weaknesses can ultimately lead under- or overreliance on models, both of which are detrimental to patient care. 

Our results demonstrate that overall model performance was good with an AUC of 0.91 (95\% CI: 0.90-0.92), similar to the FDA clearance metrics of 0.93 (95\% CI: 0.92-0.94) \cite{u.s.foodanddrugadministrationcenterfordevicesandradiologicalhealthLunitINSIGHTDBT2024}.  Model performance was robust across demographic subtypes of age, race, and ethnicity. We observed a slight, but statistically significant decrease in model performance for heterogeneously dense breasts (density C) (AUC: 0.90, 95\% CI: 0.88-0.91), and a trend of decreased performance in extremely dense breasts (density D) (AUC: 0.87, 95\% CI: 0.82-0.92). This is somewhat expected as dense breasts can mask abnormalities, even with the separation of tissues via DBT. 

When evaluating performance across imaging findings, we found that model performance was robust for exams with masses and architectural distortions, however was statistically significantly worse in exams with calcifications (AUC: 0.80, 95\% CI: 0.78-0.82) Many studies have suggested that DBT suffers from decreased sensitivity for microcalcifications \cite{spanglerDetectionClassificationCalcifications2011}, possibly related to decreased resolution of DBT and/or the visibility of microcalcifications on only a small number of tomosynthesis slices. Therefore, decreased model performance for calcifications could be considered as a limitation of DBT modality rather than the model. We also observed that architectural distortion yielded a higher number of false positives than other imaging findings. We hypothesize that this may be due to the rarity of architectural distortions compared to other findings types and its disproportionate association with malignancy. A unique feature of the BI-RADS lexicon is the various subtypes of lesions that are described; for example, a mass is described by its shape, margins and density. When evaluating performance by these features, we observe further interesting trends such as the propensity for malignant oval or circumscribed masses to have lower scores and benign irregular masses to have higher scores. We also observe that malignant exams with grouped, linear, and clustered calcifications tend to have lower scores than regional and segmental calcifications. 

When evaluating across to pathologic outcomes, we observed a statistically significant decrease in model performance for non-invasive cancers compared to invasive cancers (AUC: 0.85 vs 0.94, p<0.05). Further, the recall for non-invasive cancers was only 0.55 indicating that the model misses nearly half of non-invasive cancers We hypothesize that this is due to the tendency of non-invasive cancers to present as micro-calcifications. Nevertheless, this highlights the need for radiologists to understand the differences between machine learning metrics such as AUC, recall, and precision, as a misunderstanding of a high AUC score can lead to unintentional automation bias. Our failure analysis reveals that 44\% of missed cancers were invasive, 56\% were non-invasive, and 11\%, 61\%, 36\%, 10\% contained masses, calcifications, asymmetries, and architectural distortions, respectively. We also noted that subtypes of IDC, such as tubular, papillary, and mucinous carcinoma had higher rates of false negatives.

By examining model prediction scores at the exam level, we can detect trends that provide additional insight. For example, the number of false positives increased across screen negative, diagnostic negative, and biopsy-proven benign exams. This is not unexpected, as the degree of visualized abnormality increases between these exam types and suggests that the model is susceptible to similar false positive errors as humans. Conversely, the high number of true negatives for diagnostic negative and biopsy-proven benign cases (both of which are deemed abnormal on screening by a radiologist), suggests that the model could be beneficial in decreasing false positive recalls and biopsies. Further work is needed to assess whether a higher operating point could result in a meaningful decrease in false positive recalls and biopsies. 

\subsection{Limitations}
This study has several limitations. This is a single institution study with exams obtained across four sites. Our exams are exclusively from Hologic devices, so results may not be applicable to other scanners. All DBT exams are part of ComboHD exams, meaning that a full-field digital mammogram (FFDM), DBT, and synthetic 2D exam are obtained for each patient. Therefore, it cannot be certain that each lesion was detectable on DBT. Lastly, this model, like most other commercial AI models interprets images from only the current exam and therefore does not benefit from comparison with priors. It is possible that some lesions were apparent to the radiologist only when comparing to prior exams. Future work will include failure analysis to assess whether lesions were visible on DBT alone and whether availability of priors affects radiologist performance.

%% file: sections/5_conclusion.tex
This study represents the first thorough subgroup evaluation of a DBT model across demographic, imaging, and pathology subgroups. Our results demonstrate that model performance was robust across demographics, however, was decreased for non-invasive cancers, calcifications, and dense breasts. This highlights the need for continued vigilance in evaluation and trust of AI model performance to drive adoption and realize the promised performance gains of AI.

%% file: sections/6_acknowledgements.tex
This study was funded by Lunit, Inc. and AIM-AHEAD. We also acknowledge support from the AI Image Extraction Core, an Emory Integrated Core Facility.

%% file: sections/a1_supplement.tex
We examined the effect of various BI-RADS descriptors on model predictions, including mass shape, mass margin, calcification morphology, calcification distribution, types of asymmetries, and confidence of architectural distortion. For both masses and calcification, the most common radiologist descriptor on screening mammography was generic mass or generic calcifications without further characterization; this is often done because many radiologists prefer to under-describe lesions on screening mammography to maintain reporting flexibility on subsequent diagnostic exams. For the specific analysis of lesion features in the supplement, these exams were excluded from evaluation. 

\begin{figure}[htbp]
\centering
\includegraphics[width=0.9\linewidth]{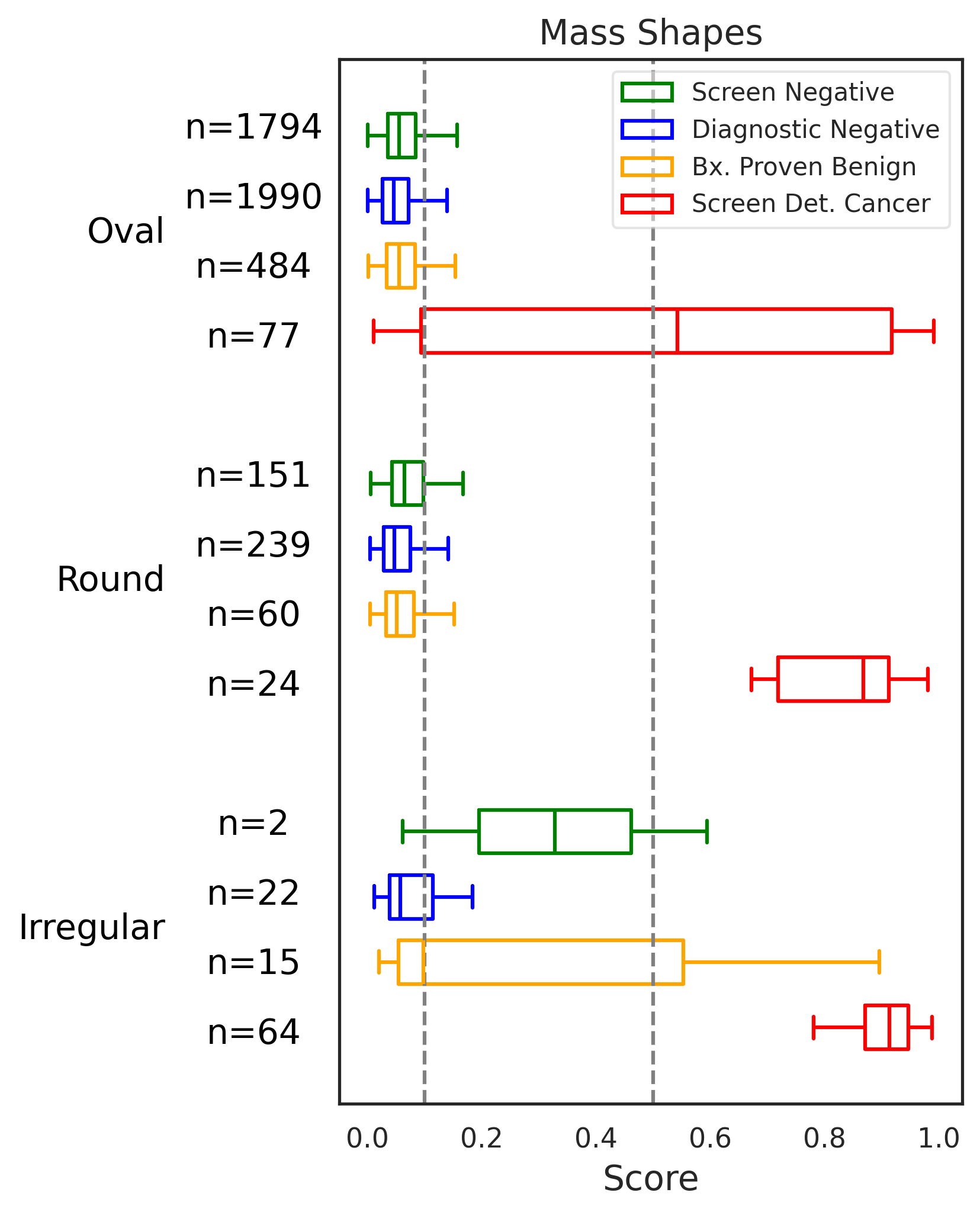}
\caption{Forest plot showing the median, 25th, and 75th percentile for model results across each BI-RADS descriptor for mass shape, further sub categorized by outcome labels. In general, the model returns higher scores for cancer cases compared to negative exams. Scores for irregular masses trended higher across all outcome types compared to round and oval masses. Of note, the median score across mass shape and outcome label is lower than the model threshold of 0.1 for oval and round masses, for negative exams.}
\label{fig:mass_shapes_by_outcome_characteristics_boxplot}
\end{figure}

\begin{figure}[htbp]
\centering
\includegraphics[width=0.9\linewidth]{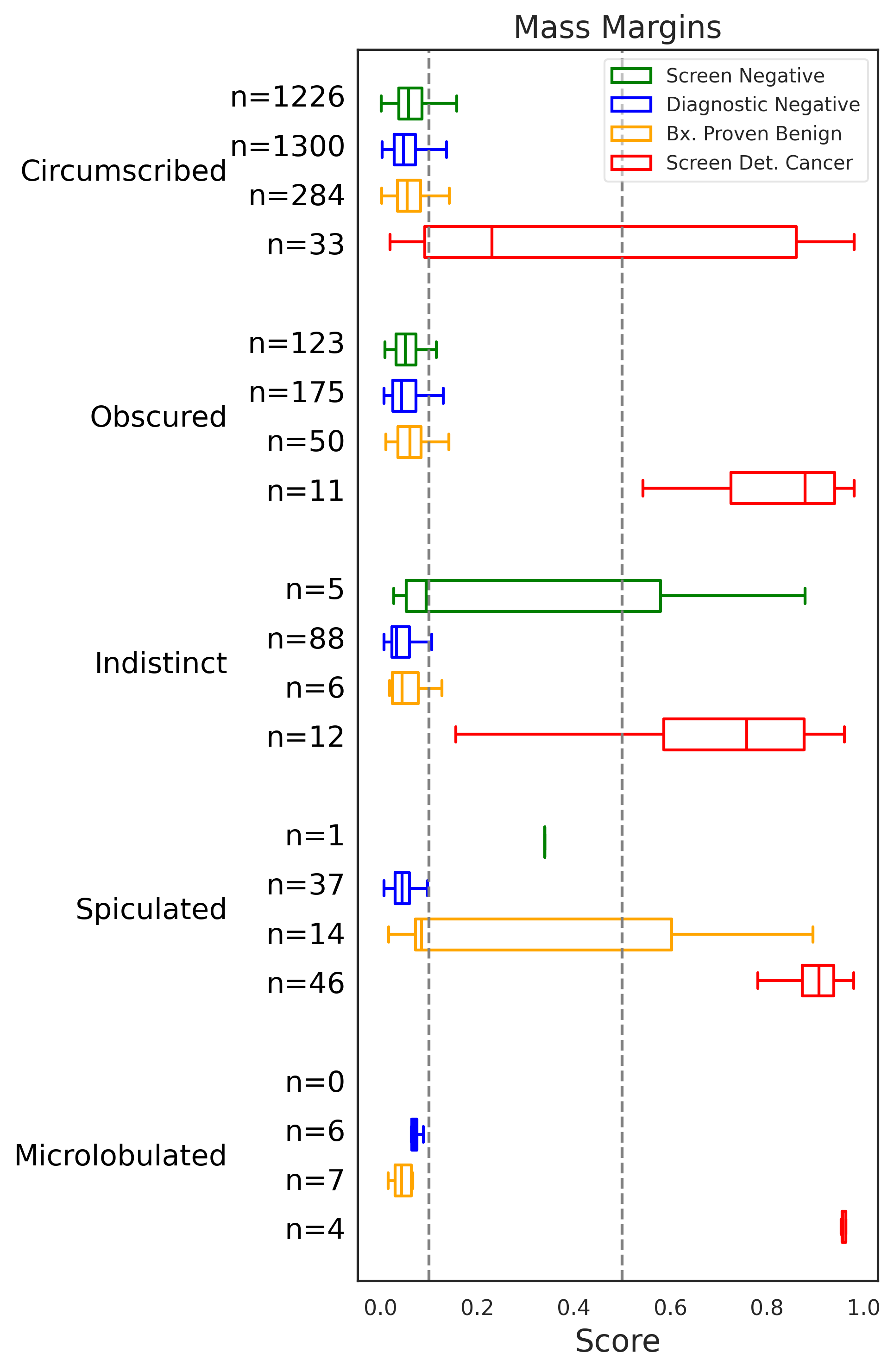}
\caption{Forest plot showing the median, 25th, and 75th percentile for model results across each BI-RADS descriptor for varied mass margins, further subcategorized by outcome labels. In general, the model returns higher scores for cancer cases compared to negative exams. Median scores for screen negative masses trended slightly higher than diagnostic negative masses across all mass margin types where there exist screen negative masses. Notably, the median score across mass margin and outcome label is lower than the model threshold of 0.1 for circumscribed and obscured mass margins, for negative exams.}
\label{fig:mass_margins_by_outcome_characteristics_boxplot}
\end{figure}

\begin{figure}[htbp]
\centering
\includegraphics[width=0.9\linewidth]{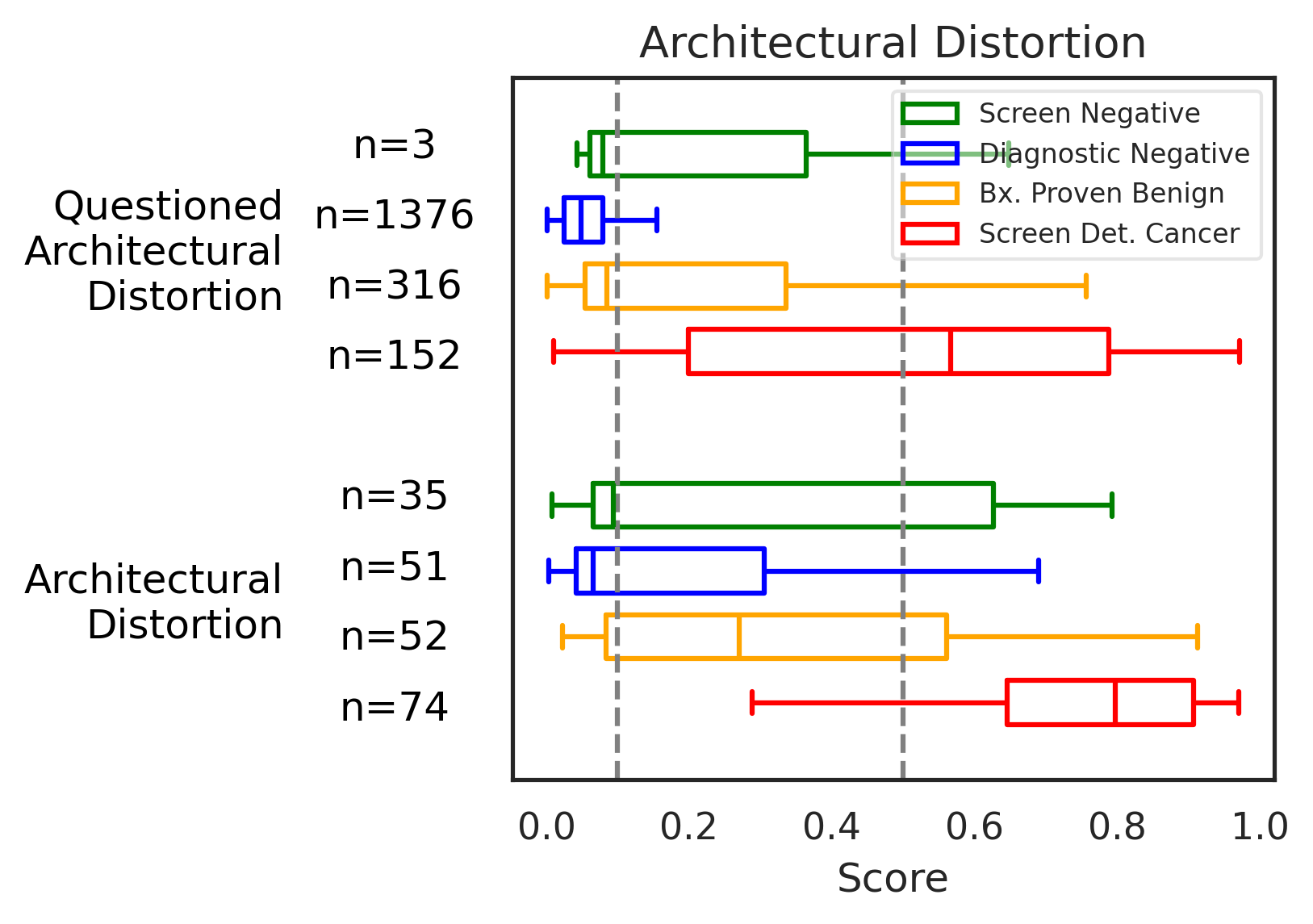}
\caption{Forest plot showing the median, 25th, and 75th percentile for model results across each BI-RADS descriptor for architectural distortion, further subcategorized by outcome labels. In general, the model returns higher scores for cancer cases compared to negative exams. Median scores for screen negative architectural distortion trended slightly higher than diagnostic negative.}
\label{fig:arch_distortion_by_outcome_characteristics_boxplot}
\end{figure}

\begin{figure}[htbp]
\centering
\includegraphics[width=0.9\linewidth]{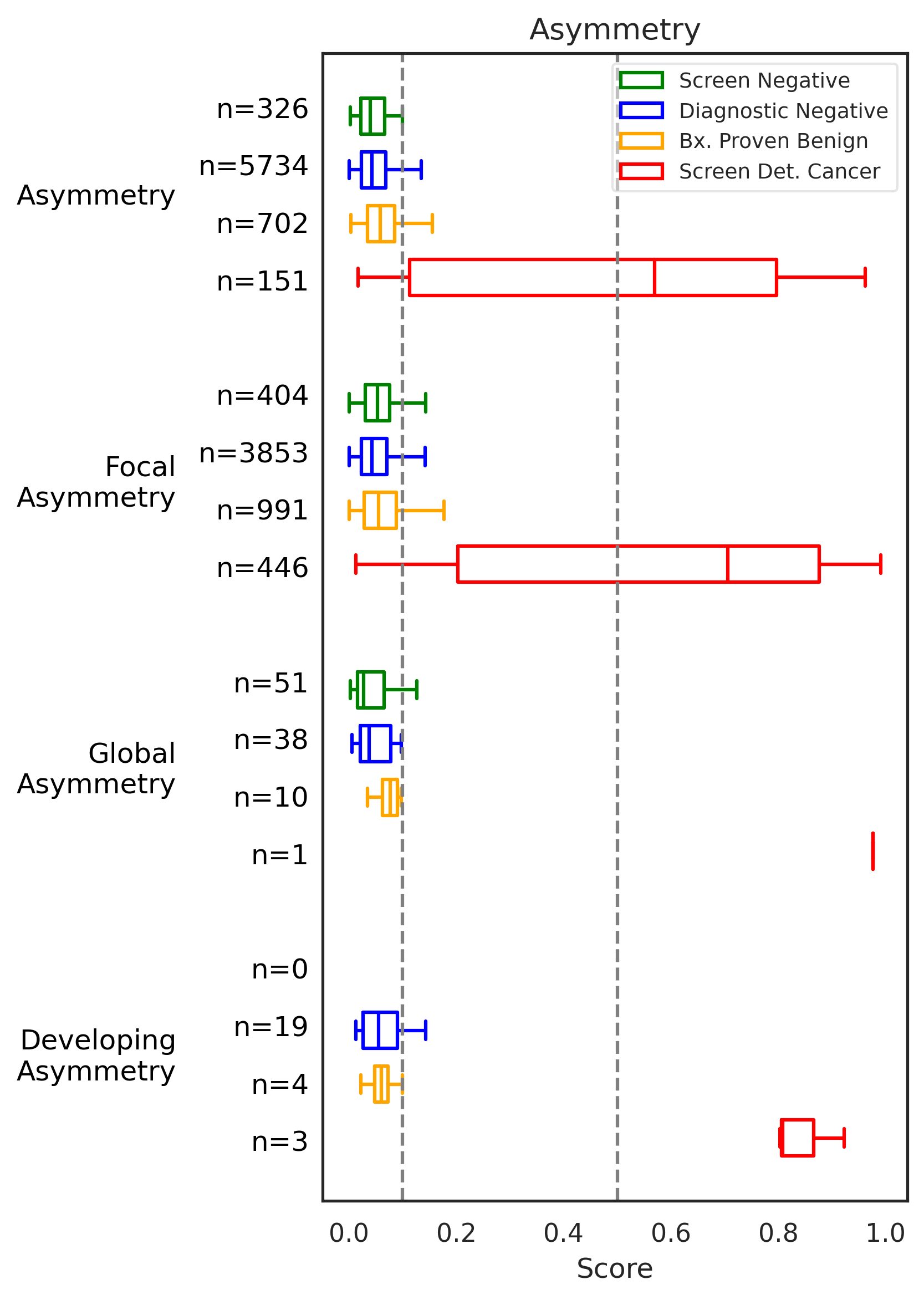}
\caption{Forest plot showing the median, 25th, and 75th percentile for model results across each BI-RADS descriptor for varied asymmetry types, further sub categorized by outcome labels. In general, the model returns higher scores for cancer cases compared to negative exams. Median scores for screen negative asymmetry trended slightly higher than diagnostic negative asymmetry for focal asymmetry. Notably, asymmetry and global asymmetry exhibited a progressive trend of increasing scores as severity of outcome increased.}
\label{fig:asymmetry_by_outcome_characteristics_boxplot}
\end{figure}

\begin{figure}[htbp]
\centering
\includegraphics[width=0.9\linewidth]{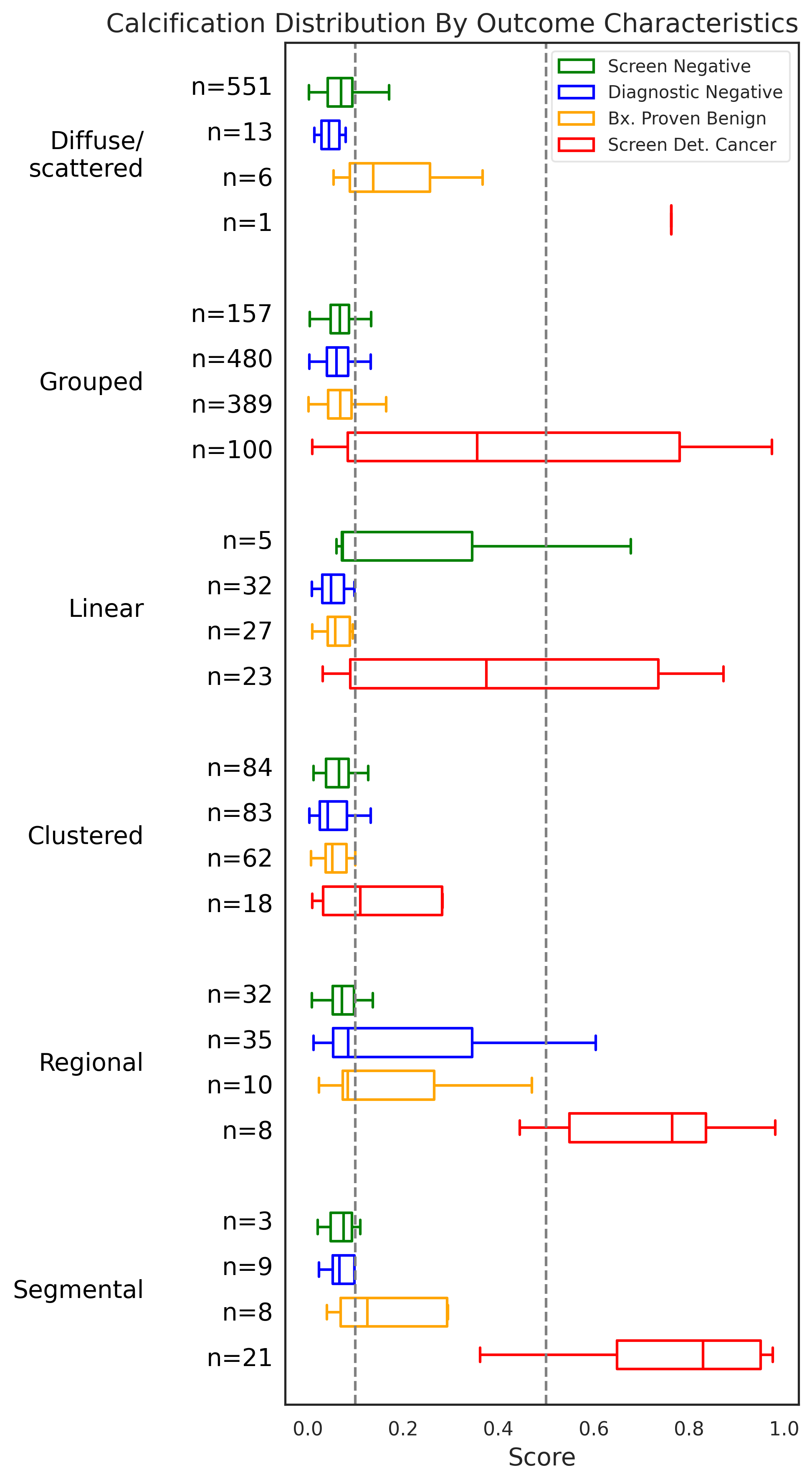}
\caption{Forest plot showing the median, 25th, and 75th percentile for model results across each BI-RADS descriptor for varied calcification distribution types, further subdivided by outcome labels. In general, the model returns higher scores for cancer cases compared to negative exams. Median scores for screen negative calcifications trended slightly higher than diagnostic negative calcifications across all calcification distribution types except for calcifications that are regionally distributed. Of note, the median score across calcification distributions and outcome label is lower than the model threshold of 0.1 for negative exams except for the confirmed benign calcifications of types diffuse/scattered and regional and diagnostic negative regional calcifications.}
\label{fig:calc_distributions_by_outcome_characteristics_boxplot}
\end{figure}

\begin{figure}[htbp]
\centering
\includegraphics[width=0.9\linewidth]{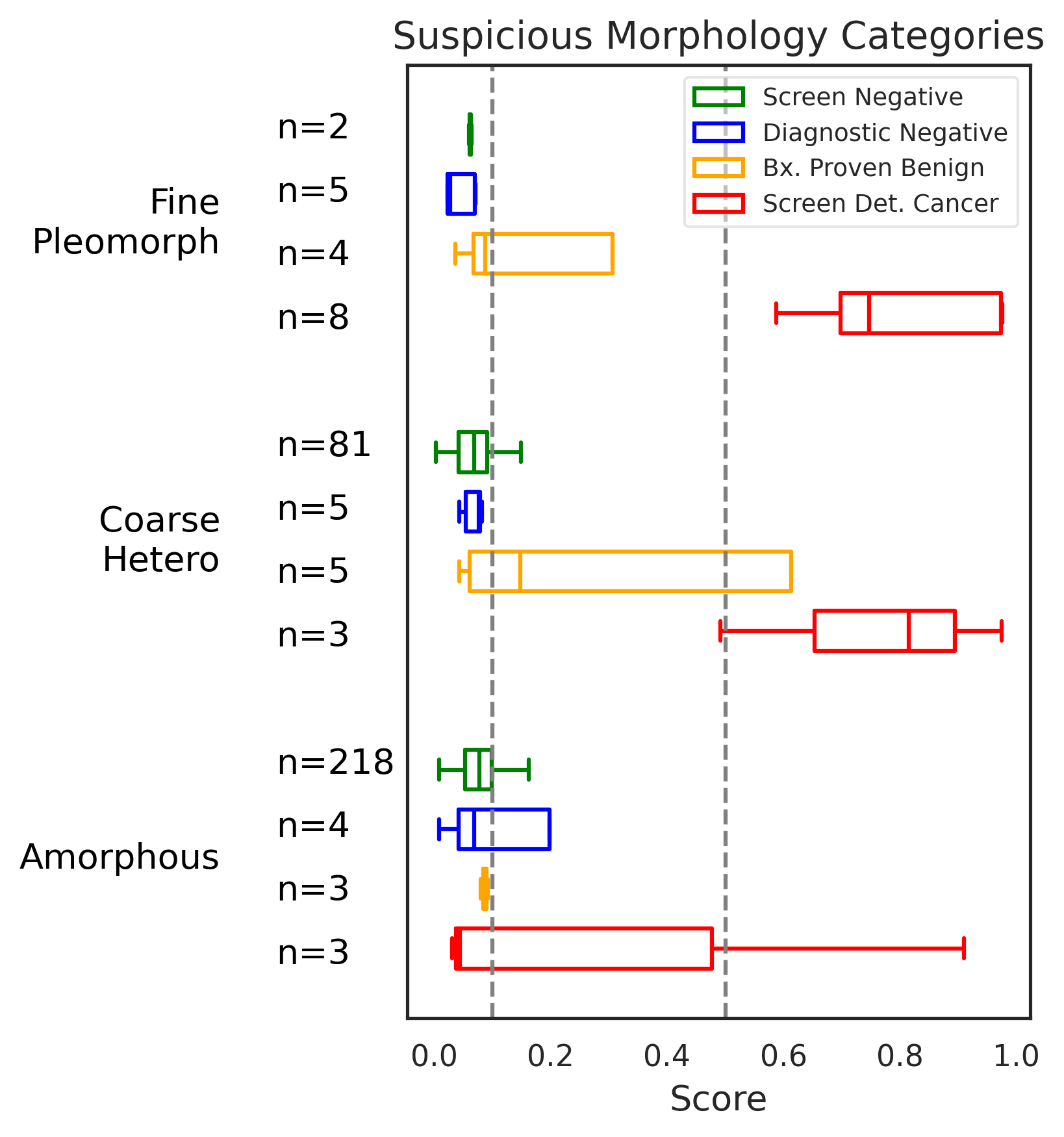}
\caption{Forest plots showing the median, 25th, and 75th percentile for model results across each BI-RADS descriptor for varied calcification morphology types, further subdivided by outcome labels. In general, the model returns higher scores for cancer cases compared to negative exams. The scores for cancerous amorphous calcifications are less than the model threshold of 0.1.}
\label{fig:calc_findings_by_outcome_characteristics_suspicious_morphology_boxplot}
\end{figure}

\begin{figure}[htbp]
\centering
\includegraphics[width=0.9\linewidth]{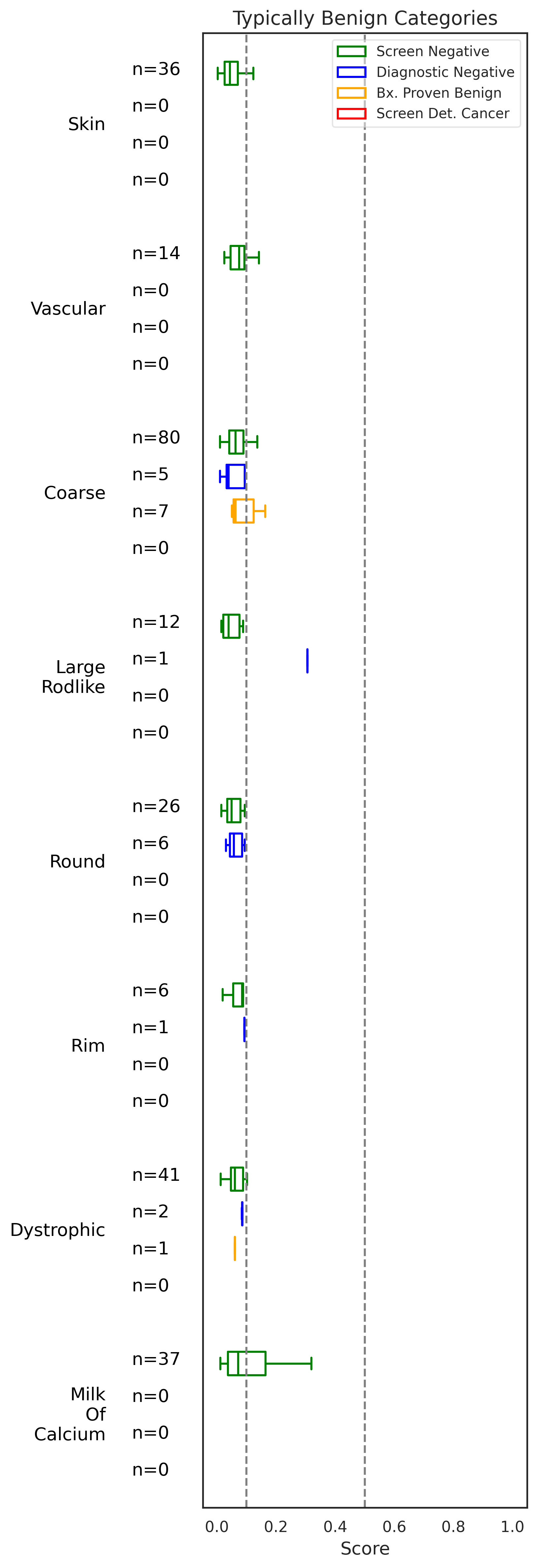}
\caption{Forest plots showing the median, 25th, and 75th percentile for model results across each BI-RADS descriptor for varied calcification morphology types, further subdivided by outcome labels. In general, the model returns higher scores for cancer cases compared to negative exams. The scores for cancerous amorphous calcifications are less than the model threshold of 0.1.}
\label{fig:calc_findings_by_outcome_characteristics_typically_benign_boxplot}
\end{figure}

\input{tables/fn_analysis_table}

%% file: tables/fn_analysis_table.tex
\begin{table*}[htbp]
\centering
\caption{Distribution of Imaging Features and False Negative Outcomes at a Threshold of 0.1. Percentages are calculated using the total number of invasive or non-invasive cancer exams as the denominator. Note that total percentages may exceed 100\% because some exams contain multiple findings in one or both breasts. Calcifications in invasive cancer exams account for 11\% of false negatives. Asymmetry identified in non-invasive cancer exams accounts for 54\% of false negatives. Additionally, 89\% of non-invasive cancer exams show calcifications.}
\label{tab:fn_analysis_table}
\begin{tabular}{rrrrr}
\toprule
\multicolumn{1}{c}{ } & \multicolumn{2}{c}{Overall} & \multicolumn{2}{c}{FN} \\
\cmidrule(l{3pt}r{3pt}){2-3} \cmidrule(l{3pt}r{3pt}){4-5}
 & Invasive
(n=914) & Non-Invasive
(n=454) & Invasive (n=160) & Non-Invasive (n=206)\\
\midrule
Mass & 233 (25\%) & 32 (7\%) & 22 (9\%) & 17 (53\%)\\
Calcification & 270 (30\%) & 406 (89\%) & 45 (17\%) & 179 (44\%)\\
Arch Distortion & 203 (22\%) & 23 (5\%) & 27 (13\%) & 9 (39\%)\\
Asymmetry & 509 (56\%) & 92 (20\%) & 87 (17\%) & 43 (47\%)\\
\bottomrule
\end{tabular}
\end{table*}